\documentclass[a4paper,11pt]{article}
\pdfoutput=1 
\usepackage{jinstpub} 
\usepackage[T1]{fontenc} 
\usepackage{appendix}
\usepackage{lineno}

\title{\boldmath Identifying Heavy-Flavor Jets Using Vectors of Locally Aggregated Descriptors}

\author[a, b]{Jana Biel\v{c}\'{i}kov\'{a},}
\author[c]{Raghav Kunnawalkam Elayavalli,}
\author[a, b, d]{Georgy Ponimatkin,\footnote{Corresponding Author}}
\author[c]{J{\"o}rn H. Putschke,}
\author[d, e]{and Josef Sivic}

\affiliation[a]{Nuclear Physics Institute of the CAS, Rez 250 68, Czech Republic}
\affiliation[b]{Czech Technical University in Prague, FNSPE, Prague 115 19, Czech Republic}
\affiliation[c]{Wayne State University, Detroit MI, USA 48201}
\affiliation[d]{Czech Institute of Informatics, Robotics and Cybernetics at the Czech Technical University in Prague}
\affiliation[e]{Inria, Ecole Normale Superieure, CNRS, PSL Research University, Paris, France}

\emailAdd{bielcikova@ujf.cas.cz}
\emailAdd{raghavke@wayne.edu}
\emailAdd{ponimatkin@ujf.cas.cz}
\emailAdd{joern.putschke@wayne.edu}
\emailAdd{josef.sivic@cvut.cz}

\abstract{Jets of collimated particles serve a multitude of purposes in high energy collisions. Recently, studies of jet interaction with the quark-gluon plasma (QGP) created in high energy heavy ion collisions are of growing interest, particularly towards understanding partonic energy loss in the QGP medium and its related modifications of the jet shower and fragmentation. Since the QGP is a colored medium, the extent of jet quenching and consequently, the transport properties of the medium are expected to be sensitive to fundamental properties of the jets such as the flavor of the parton that initiates the jet. Identifying the jet flavor enables an extraction of the mass dependence in jet-QGP interactions. We present a novel approach to tagging heavy-flavor jets at collider experiments utilizing the information contained within jet constituents via the \texttt{JetVLAD} model architecture. We show the performance of this model in proton-proton collisions at center of mass energy $\sqrt{s} = 200$ GeV as characterized by common metrics and showcase its ability to extract high purity heavy-flavor jet sample at various jet momenta and realistic production cross-sections including a brief discussion on the impact of out-of-time pile-up. Such studies open new opportunities for future high purity heavy-flavor measurements at jet energies accessible at current and future collider experiments.}


\arxivnumber{2005.018429} 

\begin{document} 
\maketitle
\flushbottom

\section{Introduction}

Jets arise from hard scattering of quarks/gluons in high energy collisions resulting in a collection of collimated particles in the detector. Jets are multi-scale objects that are sensitive to perturbative physics, such as in their production and in their parton shower or evolution~\cite{PhysRevLett.39.1436, PhysRevD.18.3320, Gribov:1972ri, Lipatov:1974qm, Dokshitzer:1977sg, Altarelli:1977zs, Catani:1996vz} and non-perturbative effects such as in hadronization~\cite{Andersson:1983ia, Webber:1983if, Wobisch:1998wt}. Recent theoretical advancements have extended quantum chromo-dynamics (QCD) calculations of the jet production cross-sections to beyond leading-order~\cite{Currie:2016bfm} and leading-log~\cite{Kang:2017frl, Dasgupta:2018nvj, Dasgupta:2020fwr} and their resummations results in predictions that reproduce trends in data over several orders of magnitude for different collision energies. In addition to being useful in extracting the strong coupling constant ($\alpha_{S}$)~\cite{Britzger:2017maj}, jets have irreversibly established themselves as key probes of QCD, searches of beyond the standard model particles and in explorations of the quark-gluon plasma (QGP) produced in relativistic heavy ion collisions. Jets produced in heavy ion collisions undergo a phenomenon called jet quenching, which manifests as energy-loss and modifications to the jet structure due to interactions with the QGP medium. Jet quenching is an important signature of the QGP and we can extract the medium transport properties by comparing data with theoretical calculations of energy loss~\cite{Burke:2013yra, Soltz:2019aea}. For more details regarding jets in heavy ion collisions, we refer the reader to these review articles from experiment~\cite{Connors:2017ptx} and theory~\cite{Qin:2015srf, Blaizot:2015lma}. Thus, in both these seemingly orthogonal areas of jet physics, the ability to identify and characterize a jet based on fundamental properties in relation to its progenitor parton such as its energy, momentum and flavor are desired. 

In this paper we focus on the topic of identifying or tagging a jet based on the flavor of the hard scattered parton. In the case of jet quenching studies, knowing the jet flavor presents an opportunity to systematically study the mass dependence of parton energy loss in the QGP. When only considering QCD radiations, the mass of the radiating particle effectively controls the phase space of the radiation as prescribed in the dead-cone effect~\cite{Dokshitzer_1991}. The dead-cone effect has been measured and studied in electron-positron collisions~\cite{Abdallah:2008ac} and recently explored in pp collisions at the Large Hadron Collider (LHC)~\cite{Maltoni:2016ays, Cunqueiro:2018jbh, Zardoshti:2020cwl}. In heavy ion collisions, the mass dependence of energy loss is still an open question with measurements at the LHC~\cite{Chatrchyan:2013exa, Sirunyan:2018jju} showing no significant differences between jets identified as heavy-flavor jets originating from a $b/c$-quark to light-flavor (gluons and $u,d,s$-quarks). It is possible that jets at the LHC (momenta $O(100-1000)$ GeV) are in high energy domain where the originating parton mass does not play a significant role in its interactions with the medium. Therefore such studies are especially important at the Relativistic Heavy Ion Collider (RHIC), where the smaller center of mass energy ($\sqrt{s} = 200$ GeV) produces jets with momenta $O(10)$ GeV which are expected to a greater extent to undergo interaction with medium and thus have enhanced sensitivity to parton flavor and mass~\cite{Kang:2018wrs, Li:2018ybp}. The smaller collision energy however makes the measurement challenging due to the significantly smaller jet production cross-section which in turn introduces a dependence on available statistics. 

We present a machine learning (ML) model which utilizes experimental information based on jet and its constituents and we identify and tag populations of light- and heavy-flavor jets with increased efficiency and purity as compared to current state of the art classifications algorithms. In contrast to light quarks, the heavy-flavor quarks are produced early in the hard scattering due to their large mass and travel a significant distance in the detector before they decay. Upon jet evolution involving fragmentation and hadronization, these massive quarks leave a characteristic experimental signature of charged particle tracks pointing back to a displaced (secondary) vertex, as opposed to the primary vertex which corresponds to the hard scattering point of interaction. These vertices can be measured by high-resolution tracking detectors. Since these displaced vertices are an important feature of heavy-flavor jets, classification algorithms predominantly take into consideration some experimental quantity related to the displaced vertex such as the distance of closest approach ($DCA$), or the secondary vertex mass amongst others~\cite{Aad:2015ydr, Chatrchyan:2012jua}. At LHC energies, it is important to note that highly-virtual gluons originating from the hard scattering could produce jets that mimic heavy-flavor jets~\cite{Ilten:2017rbd, Voutilainen:2015lqa}. This process is often treated as a part of the background since the gluon can split to a pair of heavy-flavor quarks during its evolution which could behave as a heavy-flavor jet. At the center of mass energies available at RHIC~\cite{hanseulOhSTARposter}, this gluon splitting process is significantly suppressed due to the jet kinematics and the steeply falling parton momentum spectra.   

There are two general categories of measurements involving identifying heavy-flavor jets i.e, ensemble based approaches and jet-by-jet approaches. Extracting the heavy-flavor jet fraction from an inclusive jet sample is typically done via template fits to utilizing distributions of signal and background. The latter approach of identifying jets individually by associating a light- or heavy-flavor probability is more adaptable to ML approaches. The early examples of such taggers employed boosted-decision trees (BDT) and shallow neural networks (NN) to train on a sample of signal and background jets which were subsequently applied on data after correcting for the differences between data and the simulations~\cite{Chatrchyan:2012jua, Jung:2016yzu, Aad:2015ydr}. Current state of the art studies and measurements at the LHC have expanded to include deep, convolutional and recurrent networks~\cite{PhysRevD.94.112002, Sirunyan:2019wwa, ATL-PHYS-PUB-2017-003, ATL-PHYS-PUB-2017-013}. The classification procedure where the networks were trained on MC with associated signal/background labels is commonly known as supervised training and is dependent to an extent on the MC accurately representing data. 

Since jets are essentially collections of objects (tracks/towers in experiment and particles in MC), a majority of the high performing heavy-flavor classification models currently used in experiment utilize information contained within these jet constituents. Experiments with charged particle and vertex tracking detectors with high pointing resolution, O(10-100$\mu m$), can associate tracks originating from different vertices. Providing the jet constituents ($4-$momenta and vertex information) to a sufficiently complicated model should effectively include all available physics required to distinguish between heavy- and light-flavor jets. We introduce a model that utilizes these jet constituents and study the performance in detail for jets of varying momenta and for varying categories of inputs. The rest of the paper is organized as follows. The MC samples are outlined in section~\ref{sec:dataset} along with a discussion of the different inputs types to the classification model. The \texttt{JetVLAD} model architecture is presented in section~\ref{sec:model} and we present a discussion of performance metrics that are studied in this paper in section~\ref{sec:metrics}. We discuss the results for RHIC energies in section~\ref{sec:results} and conclude our study with an outlook focusing on applicability in current and future experiments in section~\ref{sec:conclusions}.  

\section{Datasets And Inputs}
\label{sec:dataset}

We use \textsc{PYTHIA 8.235}~\cite{Sjostrand:2014zea} to generate di-jet events in proton-proton (pp) collisions at $\sqrt{s} = 200$ GeV. In order to maximize the classification performance, we name two classes of jets as light (originating from gluon, $u, d, s$ quarks) and heavy-flavor ($c, b$ quarks). Flavor labelling is done by a requiring the initiating parton to be contained with the jet radius. Recently, the heavy flavor tagging community has started to look at jet-flavor association by utilizing reconstructed mesons from heavy flavor quark such as $\rm{D_{0}}$~\cite{ALICE-PUBLIC-2020-002}. We will explore this style of identification/tagging in an upcoming publication but for the purposes of this paper, we utilize the hard-scattered quark to jet matching for defining our jet classes. To compare the effect of the production cross-section, we produce two sets of samples which are labeled as follows: 
\begin{itemize}
    \item Cross-section weighted   
    \item Balanced - 50\% light, 25\% $c$-jet and 25\% $b$-jet.
\end{itemize}
Particle decays in \textsc{PYTHIA} along $x-y$ and $z$ are limited to maximum distances to $2000$ mm and $600$ mm, respectively. For each dataset, we generate both light- and heavy-flavor di-jet events with the invariant $\hat{p_{T}}$ corresponding to $[3-12], [8-17], [13-22], [18-27]$ and $[23-42]$ GeV/$c$. The overlap in the upper and lower limits is to maximize statistics when combining the datasets together\footnote{The samples used in this study are made available along with all the necessary software tool-kits for processing and training/testing upon publication}. The datasets are split into $80:10:10$ for training, testing and validation covering a total of 2 million events for the balanced sample and 4 million events for the cross-section weighted sample.  

In order to simulate particle interaction with the detector, we apply a fast simulation (Fast-Sim) of the STAR detector~\cite{star_detector}. The Fast-Sim framework includes a parametrization of charged particle tracking efficiency, momentum resolution smearing and secondary vertex $DCA$ smearing according to the STAR Time Projection Chamber (TPC)~\cite{Anderson:2003ur} and Heavy-Flavor Tracker (HFT)~\cite{QIU20141141}, respectively. The Fast-Sim procedure is outlined in greater detail in Appendix~\ref{app:fastsmear}. Post smearing, we reconstruct jets from all smeared charged particles using the anti$-k_{t}$ reconstruction algorithm~\cite{Cacciari:2008gp} as implemented in FastJet~\cite{Cacciari:2011ma} with a jet resolution parameter $R=0.4$. The charged particles which are associated to the jet are then further considered for classification procedure. These jets are often referred to as charged-jets and future extensions of this study will consider both charged and neutral components of a jet, taking into account the energy depositions recorded by both electromagnetic and hadronic calorimeters.  

\begin{figure}
    \centering
    \includegraphics[width=0.49\textwidth]{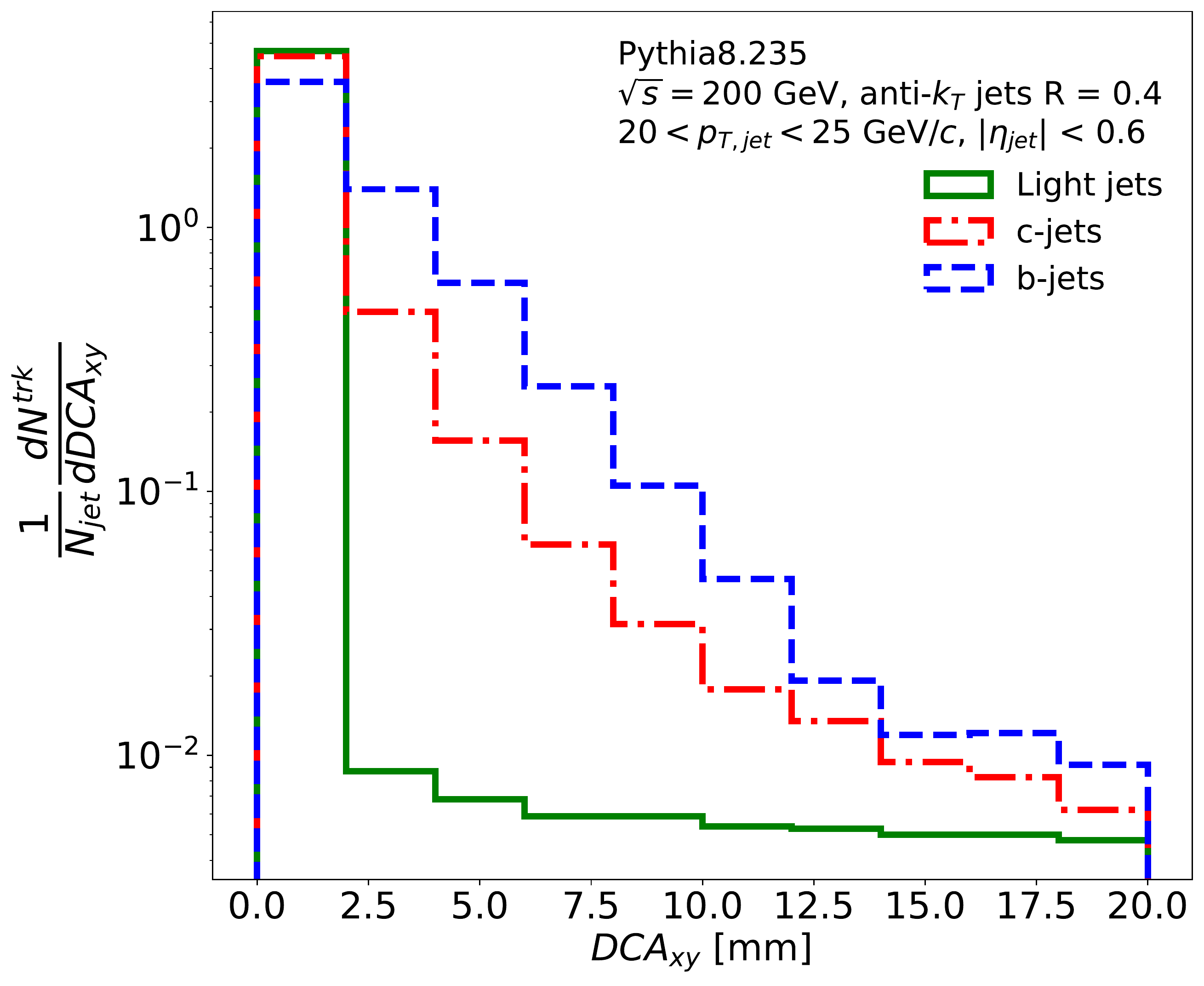}
    \includegraphics[width=0.49\textwidth]{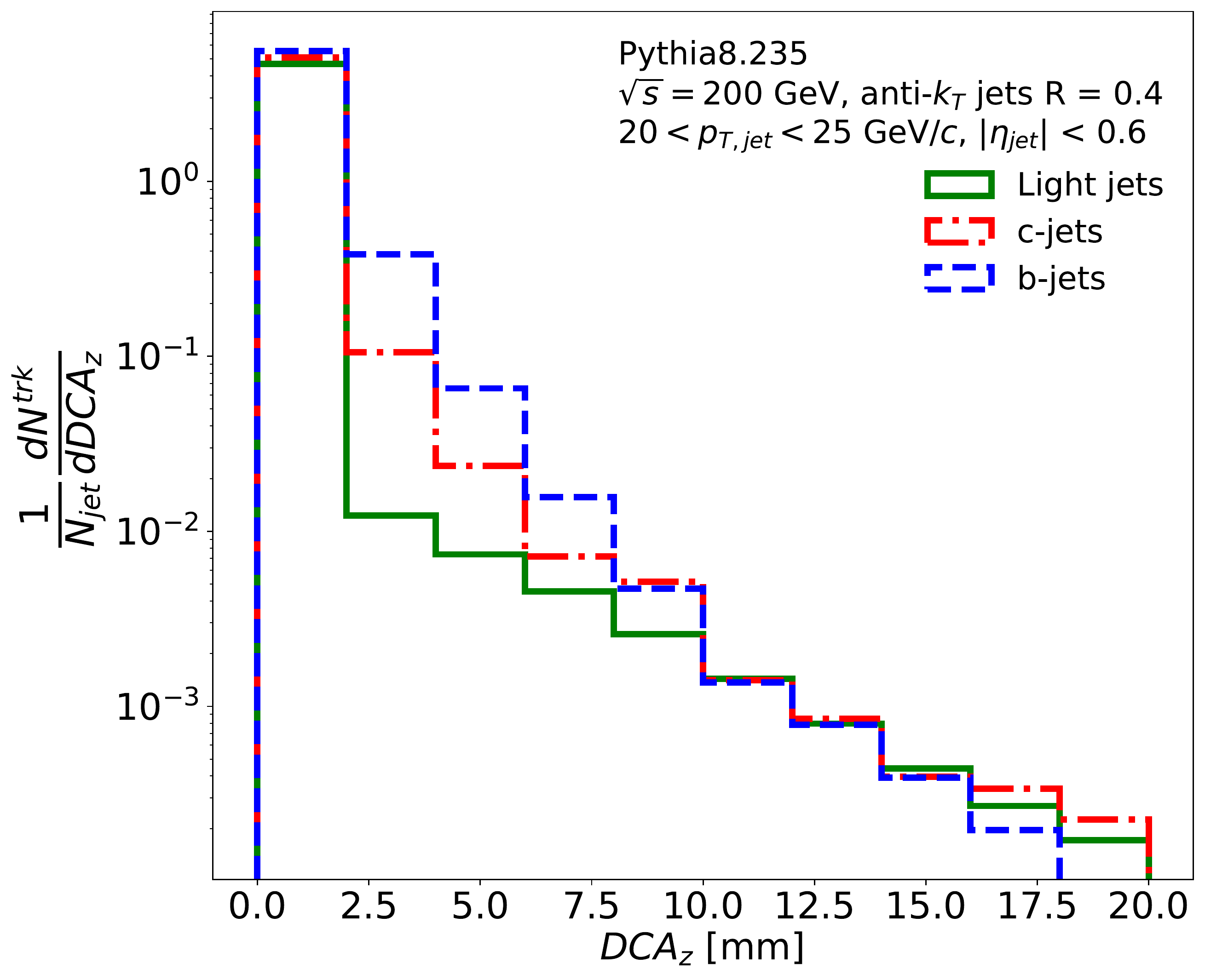}
    \includegraphics[width=0.49\textwidth]{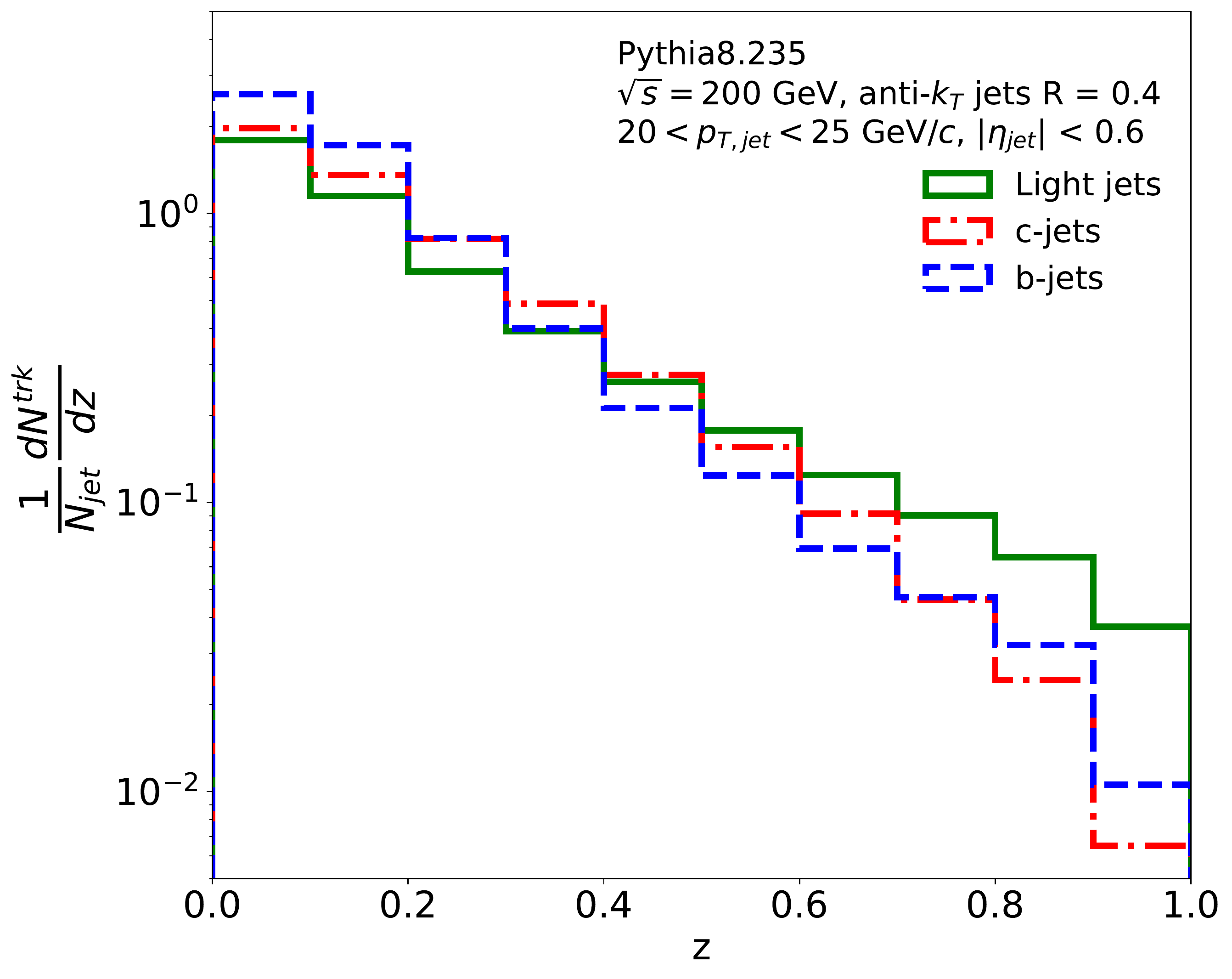}
    \includegraphics[width=0.49\textwidth]{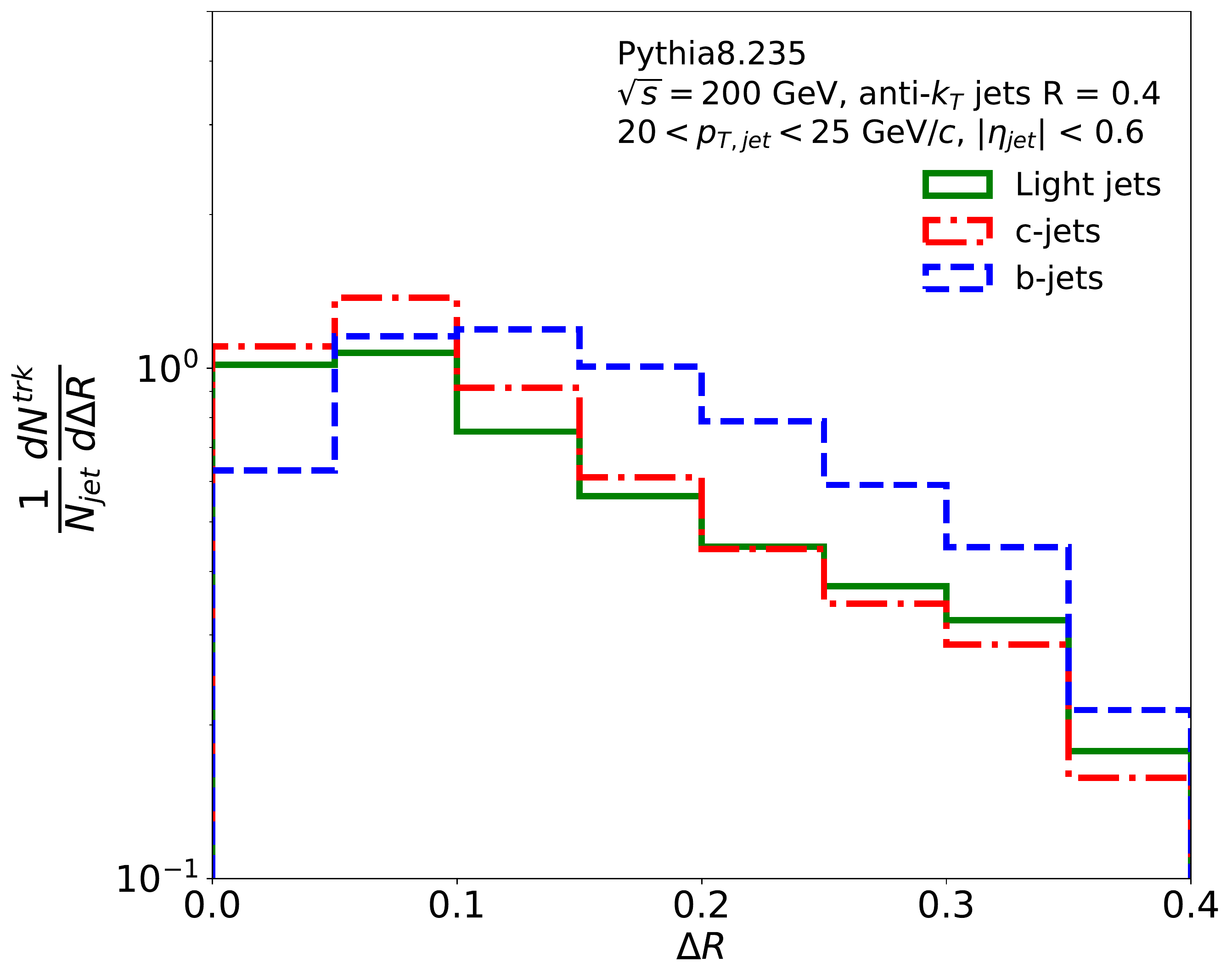}
    \caption{Distributions of a subset of input variables $DCA_{xy}$ (top left), $DCA_{z}$ (top right), $z$ (bottom left) and $\Delta R$ (bottom right) for light (black), c (red) and b-jets (blue) as generated in \sc{PYTHIA}. (color online)}
    \label{fig:inputComp}
\end{figure}

We separate the inputs into three different types including: tracking, fragmentation and secondary vertices as shown and defined in Tab.~\ref{tab:inputs}. For each of these classes, the inputs correspond to the values mentioned for each track in the jet. We do not take into account the particle ID or mass which are left for future studies. In addition to track kinematic variables, we also include one high level observable related to jet fragmentation such as the $z\Delta R^{2}$, a quantity related to the jet mass when summed over all particles~\cite{Kang:2018qra}. Furthermore, by comparing the performance of the model with tracking and vertexing or tracking and fragmentation as inputs, one can estimate the impact such information will have on the tagging of heavy-flavor jets. This can help to clarify detector performance needed in future experiments or in upgrading current experiments.

\begin{table}[h]
\centering
\begin{tabular}{| c | c | c |}
\hline
 Type & Inputs &  Definition\\ [0.5ex]
 \hline\hline
 Tracking & $p_{T}$  &  Transverse momentum in the $x-y$ plane \\
   & $\eta$  &  pseudorapidity \\
   & $\phi$  &  azimuthal angle \\
   \hline 
 Fragmentation & $z$  &  momentum fraction $\frac{p^{track}_{T}}{p^{jet}_{T}}$ \\
  & $\Delta R$  &  distance between track and jet axis  $\sqrt{\Delta \phi^{2} + \Delta \eta^{2}}$ \\
   & $z \Delta R^{2}$  & higher level feature \\
   \hline 
 Secondary Vertex & $DCA_{xy}$ & Distance of closest approach in $x-y/$ \\
  & $DCA_{z}$ & Distance between primary and secondary vertex in the $z$ axis \\
 \hline 
\end{tabular}
\caption{Input variables and their types and definitions utilized in the classification.}
\label{tab:inputs}
\end{table}

The input distributions for $20 < p_{T} < 25$ GeV/$c$ jets are shown in Fig.\ref{fig:inputComp}, $DCA_{xy}$ (top left), $DCA_{z}$ (top right), $z$ (bottom left) and $\Delta R$ (bottom right) for light jets in the green solid lines and heavy-flavor separated into $b$ and $c-$jets in the blue dashed and red dot-dashed lines, respectively. As the mass of the originating parton increases, we find jets to have a larger width in the secondary vertex and distinctive shifts in the opening angles and fragmentation. For the purposes of the training, we combine both $b$ and $c-$jets into a single class called heavy-flavor jets. 

The MC sample utilized for training including the Fast-Sim procedure, has been shown to be comparable to the STAR detector performance in previous publications~\cite{PhysRevC.102.054913, Adam:2020kug}. Since the training sample is at the detector-level (i.e after Fast-Sim), any unfolding corrections for detector effects/uncertainties are expected to be done post classification. Given that the tracking efficiency within STAR is $\pm 4\%$~\cite{Adam:2020kug}, we find no significant difference to the jet energy scale/resolution in the light- and heavy-flavor jet samples. As such, in the rest of the paper we will discuss jets (and their momenta) at the detector-level. 

Before we train our model, variables are normalized to their mean to ensure uniformity amongst datasets, which is a necessary step in ML often referred to as pre-processing. 

\section{\texttt{JetVLAD} - Model}
\label{sec:model}
There are several different ways of representing jets for machine learning tasks~\cite{Larkoski:2017jix}. Two of these are the graph-based or the set-based approach. The former comes from the fact that we can traverse back through the jet clustering history to recreate in some metric, the original parton shower. While such a graphical representation might be considered a bonus for jet flavor tagging, it has disadvantages, namely that such ordering is clearly dependent on the underlying model of parton shower evolution and also factors in the specific algorithm utilized for the de-clustering. This might lead to a wider domain gap between simulated jets from MC which are used for the training of a machine learning algorithm and experimentally reconstructed jets in data.
 
The set based approach utilizes a simpler view, where we can consider a jet as a set or collection of particles. Such approaches could also be dependent on the type of algorithm used for jet finding but those effects are typically small. As an appreciable consequence, this approach has a smaller domain gap due to an easier theoretical description of jets at this level and it has less model dependence than the graph based description. Another benefit of this approach is that primary jet finding is predominantly done using anti-$k_{t}$ jet clustering algorithm and such jets do not have a physical clustering tree (in comparison to the $k_t$~\cite{Catani:1993hr} and C/A~\cite{Dokshitzer:1997in} algorithms). 
 
We begin description of our model by formalizing the dataset notation. We are given a jet $\mathcal{J}$ composed of a set of particles which are created by an action of the anti-$k_t$ jet clustering algorithm, i.e.
\begin{equation}
\label{eq:jet}
    \mathcal{J} = \left\{(p_{T,i}, \eta_{i}, \phi_{i}, \dots)\right\}^{n}_{i = 1},
\end{equation}
where $n$ corresponds to the total number of jet constituents. In recent publications, set represented jets were classified using Recurrent Neural Network (RNN) models~\cite{Andreassen:2018apy}, where an artificial ordering in $p_T$ or vertex distance was introduced. To overcome such an arbitrary ordering, one might use an aggregation method, that will allow us to characterize a set of inputs into a fixed-length feature vector. An analogous situation is found in computer vision with respect to the procedure of place recognition, where one wants to recognize a landmark (say the Eiffel tower). In this case we often deal with a variable size set of feature descriptors extracted from the image. For example, depending on the place where the photo was taken, we might have variable amount of trees, cars and other background objects. 

The \texttt{NetVLAD} is an adaptive pooling layer that takes a set of feature descriptors as an input and returns a fixed-length feature vector that characterizes each set~\cite{Arandjelovic18, DBLP:journals/corr/MiechLS17}. While in computer vision one needs a feature extractor that yields meaningful descriptors, in physics, observables already hold rich information regarding the jet shower. Thus we can omit the feature extractor stage and use tracks belonging to a jet which we call henceforth as particle descriptors. 
The ordering of the particle descriptors in the input was varied in a randomized fashion with no particular ordering having an impact on the classification performance. 

Given a jet $\mathcal{J}$ with $n$ tracks, each of which is represented by a $d$-dimensional particle descriptor, as described in equation~\ref{eq:jet}, we define $k$ clusters in the input space of the model, where each cluster is represented by parameter vectors {$\mathbf{w}_k$, $\mathbf{c}_k$} and scalar $b_k$ that are learnt from data. The output of \texttt{NetVLAD} layer is a $d\times k$-dimensional matrix, whose elements are given by 
\begin{equation}
\label{eq:vector}
    \mathbf{V}_{j,k} = \sum_{i = 1}^n \frac{e^{\mathbf{w}^T_k\mathbf{x}_i + b_k}}{\sum_{k'} e^{\mathbf{w}^T_{k'}\mathbf{x}_i + b_{k'}}}(\mathbf{x}_{i,j} - \mathbf{c}_{k,j}).
\end{equation}
Here $\mathbf{x}_{i,j}$ is a $j$-th element of the $i$-th particle descriptor and $\mathbf{c}_{k,j}$ is the $j$-th element of the $k$-th cluster center vector. This matrix is then $L^2$ normalized column-wise, transformed into a vector and then again $L^2$ normalized. \texttt{NetVLAD} hence summarizes a set of particle descriptors into one fixed-length feature vector, that is then fed into the standard feed-forward neural network. Please note that the vectors $\mathbf{w_k}$, $\mathbf{c_k}$ and scalar $b_k$ for each of cluster are parameters of this \texttt{NetVLAD} layer and learnt from data (together with other parameters of the model) in a discriminating manner using back-propagation as described below.

We chose our network architecture to mimic the ResNet model family~\cite{DBLP:journals/corr/HeZRS15}, by utilizing residual blocks witch batch normalization, in order to simplify the learning problem. Width of our model was chosen to be the same as the output of \texttt{NetVLAD} layer. We also utilize DropOut method~\cite{10.5555/2627435.2670313} in order to increase generalization of our chosen model. Our total architecture thus can be written as
\begin{equation}
    \texttt{JetVLAD} = \texttt{NetVLAD}(N_{c}) \to D\times[\text{ResidualBlock}] \to \text{Softmax},
\end{equation}
where $N_{c}$ is the number of clusters and $D$ is the depth. We train our model using momentum stochastic gradient descent (SGD)~\cite{DBLP:journals/corr/LoshchilovH16a} with cosine annealing and a warm restart. We chose learning rate of $0.013$, and annealing parameters $T_0 = 1$ and $T_{\text{mult}} = 3$ by utilizing random grid search. The model was trained for maximum of $2000$ epochs, with early stopping criterion of $10$ epochs used. We also found that number of clusters $N_{c} = 33$ and depth $D = 4$ were good set of hyper-parameters. 

\section{Classification Performance Metrics}
\label{sec:metrics}
In order to evaluate performance of the model one has to choose a meaningful set of metrics that will quantify key aspects of model performance. The first metric is called efficiency, in physics or true positive rate (TPR), in machine learning/computer vision and it is defined as 
\begin{equation}
    \textrm{TPR} = \frac{\textrm{TP}}{\textrm{P}}.
\end{equation}
Here $\textrm{TP}$ is a number of positively identified heavy-flavor jets and $\textrm{P}$ is a total number of heavy-flavor jets in the testing sample. Hence, this metric tells us the fraction of the signal that the model will extract from the sample. 

Next metric that is closely related to the efficiency is mis-identification probability, in physics or false positive rate (FPR), in machine learning/computer vision,  
\begin{equation}
    \textrm{FPR} = \frac{\rm{FP}}{\rm{N}}.
\end{equation}
Here $\textrm{FP}$ is a number of false-positive samples identified in the testing sample and $\textrm{N}$ is a total number of background objects in the testing sample. This metric quantifies the amount of background that still persists in the signal post classification. Another related metric is background rejection (REJ), which has no analogies in machine learning literature, and is given by
\begin{equation}
\textrm{REJ} = \frac{1}{\textrm{FPR}}.
\end{equation}
This determines how much of the true background will be rejected per one false-positive detection. It is a useful quantity particularly for heavy-flavor jet classification where the signal is two orders of magnitude smaller than the background due to the difference in the production cross-sections. 

Last relevant metric is purity, in physics or precision, in machine learning and it is given by
\begin{equation}
    \textrm{PREC} = \frac{\textrm{TP}}{\textrm{TP} + \textrm{FP}},
\end{equation}
where $\textrm{TP}$ is a number of true positive objects found in the testing sample and $\textrm{FP}$ is a number of false positive objects found in the testing sample. As its name suggests, this metric tells us the extent of  contamination in the signal with false-positive objects. Summary of all the metrics and their definitions are given in Tab.~\ref{tab:metrics}.

\begin{table}[h]
\centering
\begin{tabular}{| c | c | c |}
\hline
 Physics & Machine Learning &  Definition\\ [0.5ex]
 \hline\hline
 Tagging Efficiency & True Positive Rate (TPR)/Recall  &  $\textrm{TPR} = \frac{\textrm{TP}}{\textrm{P}}$ \\
 Misidentification Prob. & False Positive Rate (FPR)  &   $\textrm{FPR} = \frac{FP}{N}$  \\
 Background Rejection &   &   $\textrm{REJ} = \frac{1}{\textrm{FPR}}$  \\
 Signal Purity & Precision  &  $\textrm{PREC} = \frac{\textrm{TP}}{\textrm{TP} + \textrm{FP}}$  \\
 \hline
\end{tabular}
\caption{Classification metrics used in physics and machine learning.}
\label{tab:metrics}
\end{table}

\section{Sensitivity To Hyper-Parameters And Model Uncertainties}
\label{sec:ablationtests}

\begin{figure}
    \centering
    \includegraphics[width=0.49\textwidth]{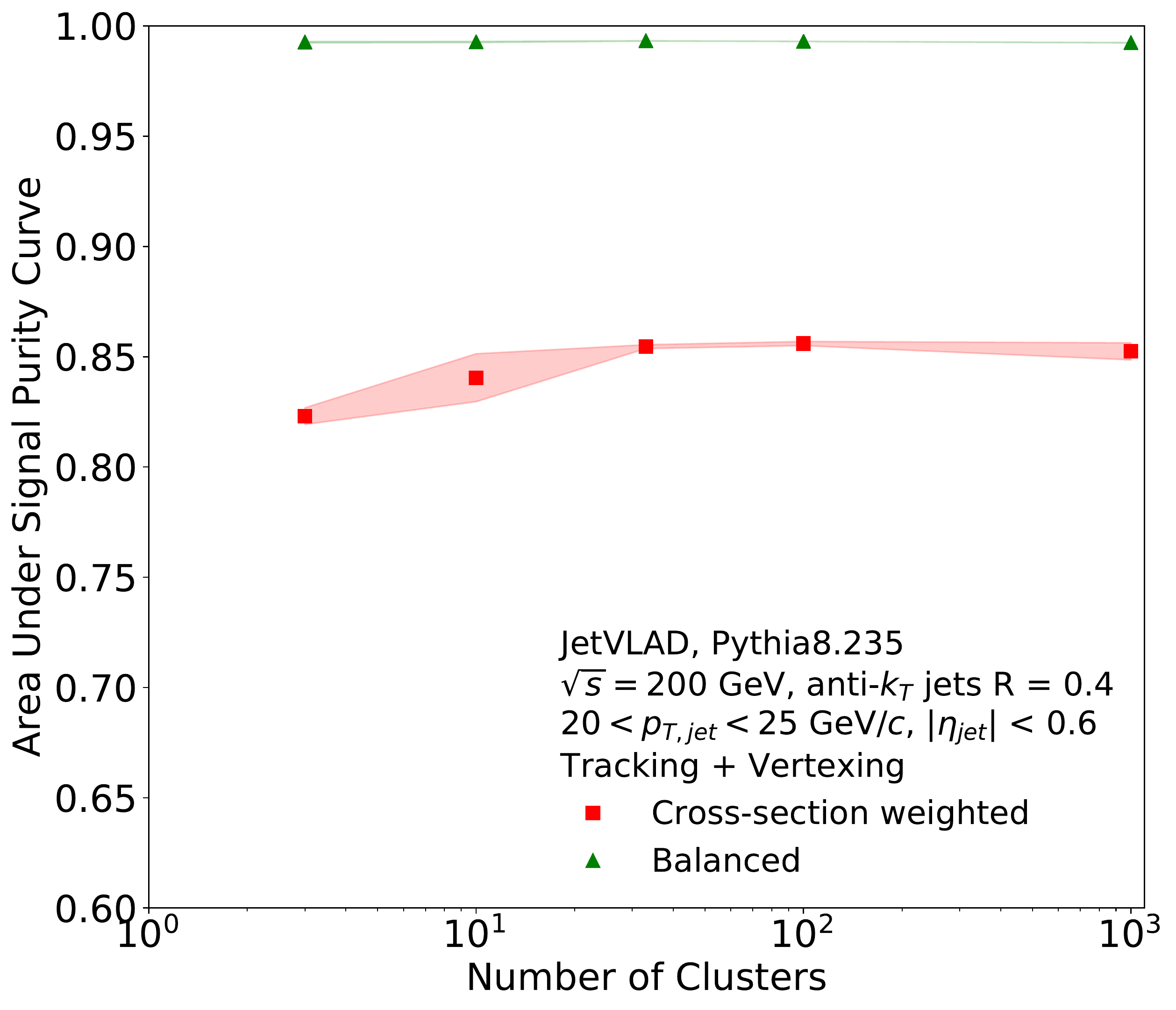}
    \includegraphics[width=0.49\textwidth]{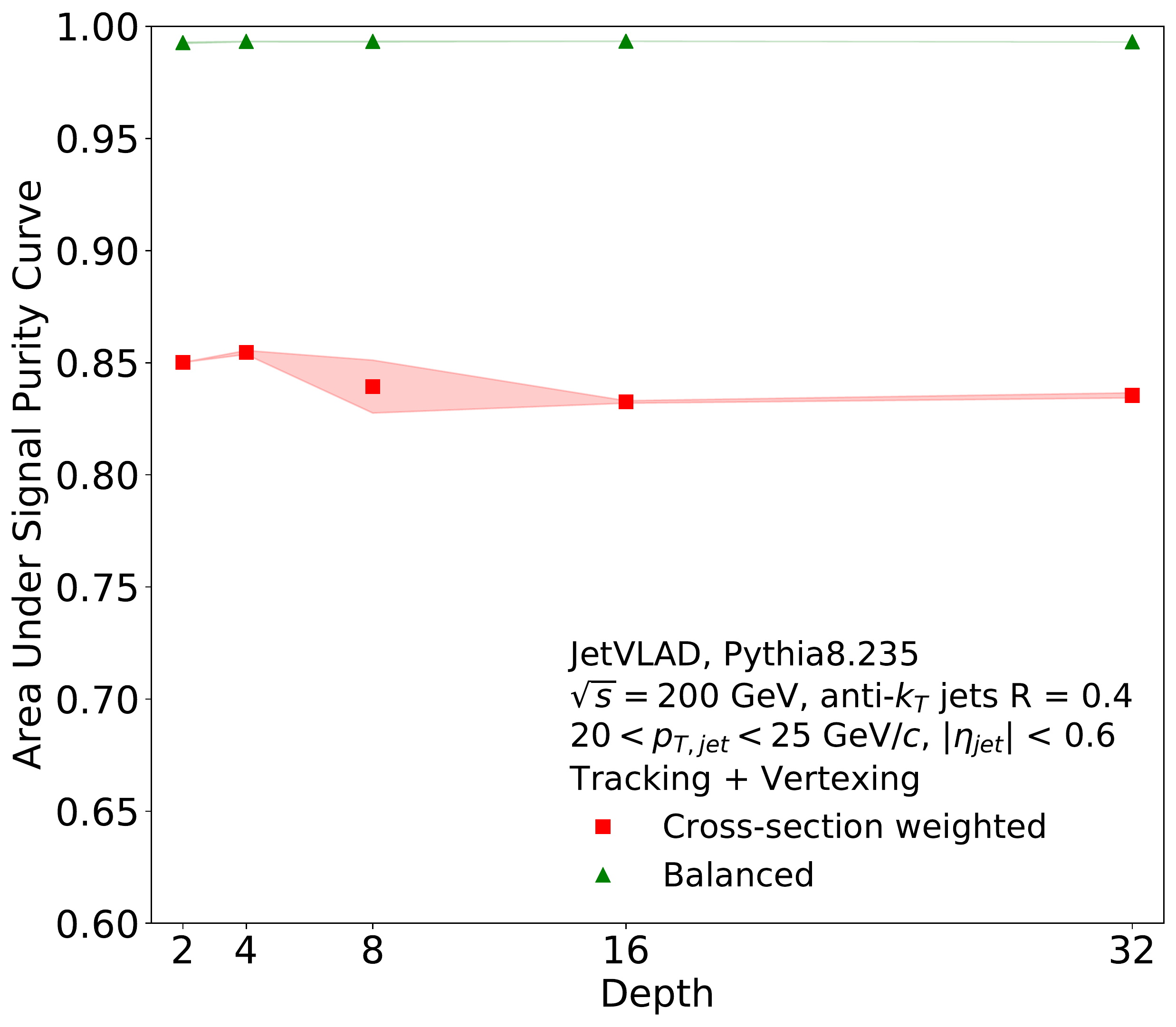}
    \caption{Hyper-parameter scans for \texttt{NetVLAD} layer showing the performance for heavy-flavor identification purity as we vary the number of clusters (left) and the model depth (right).}
    \label{fig:ablationtesteff}
\end{figure}

In order to study the dynamics of our model, we run a study to quantify effects of hyper-parameter change on the model performance. In each run, we fixed one of the parameters (for example, number of clusters) and varied the other one (model depth). Results for cluster scan, during which we fixed depth of the model $D = 4$, and varied the number of clusters can be seen in the left panel of Fig.\ref{fig:ablationtesteff}. The area under the purity curve is used as a metric of performance and we see no significant improvement beyond a total of $33$ clusters. Similarly, the depth hyper-parameter sensitivity study shown in the right panel Fig.\ref{fig:ablationtesteff} during which we fixed $N_{c} = 33$. 

Estimating systematic uncertainties of a deep neural networks is a rather recent development with different ideas~\cite{DBLP:journals/corr/abs-1907-06890, Englert:2018cfo, Kasieczka:2020vlh}. Given that the model hyper-parameters were fixed for optimal performance, modifying those is not an appropriate way to truly estimate the uncertainties inherent in the model. Given a fixed representation of \texttt{JetVLAD} used in this study, we find a total of $111608$ trainable parameters. Each of these parameters or weights are initially randomized and are later fixed during the training. We further randomize these input weights with three iterations of the model training and the different results are taken as a systematic variation of the classification and show in the corresponding shaded regions with their average taken as the central value.  

\section{Heavy-Flavor Jet Tagging At RHIC Energies}
\label{sec:results}

\begin{figure}
    \centering
    \includegraphics[width=0.49\textwidth]{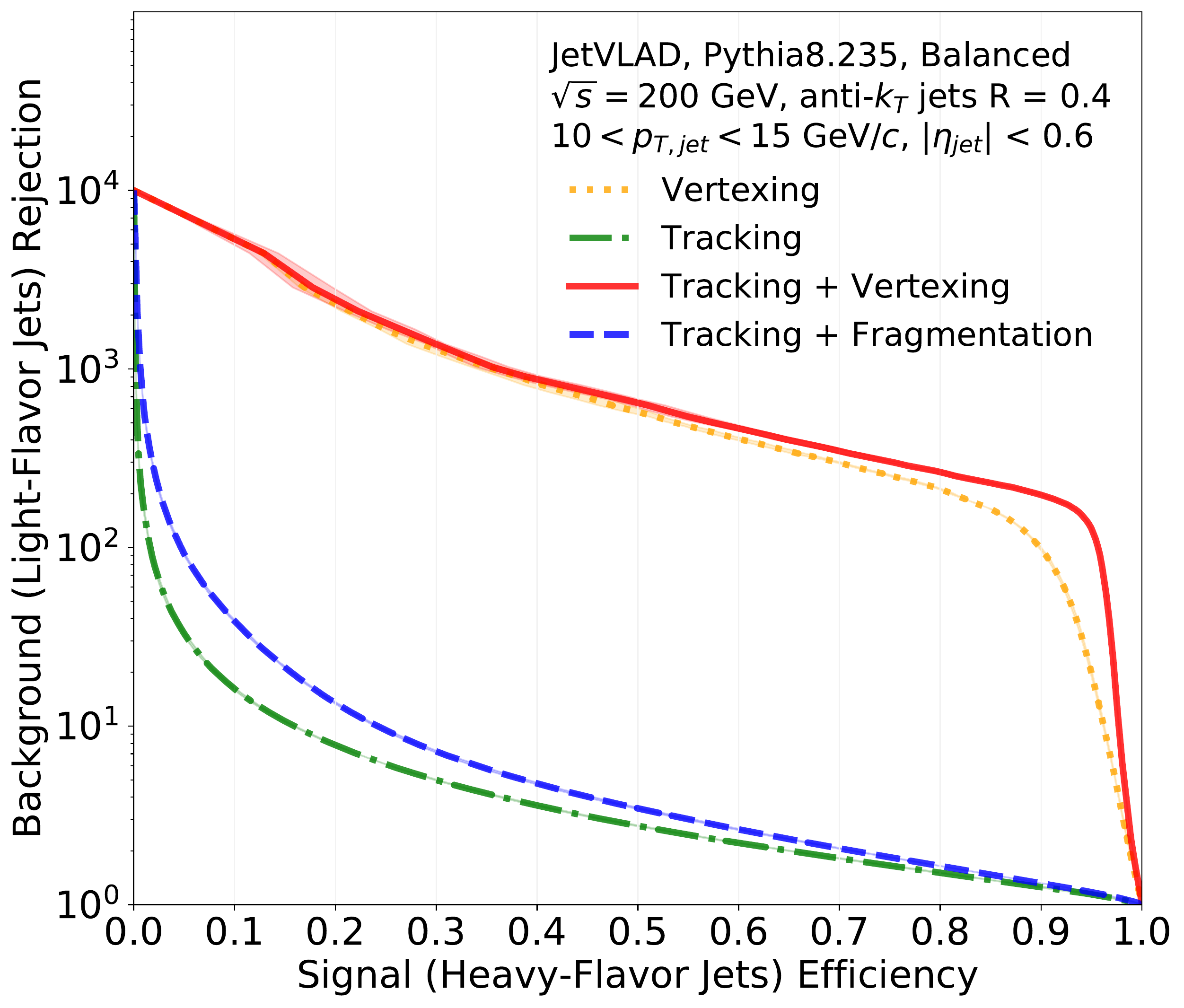}
    \includegraphics[width=0.49\textwidth]{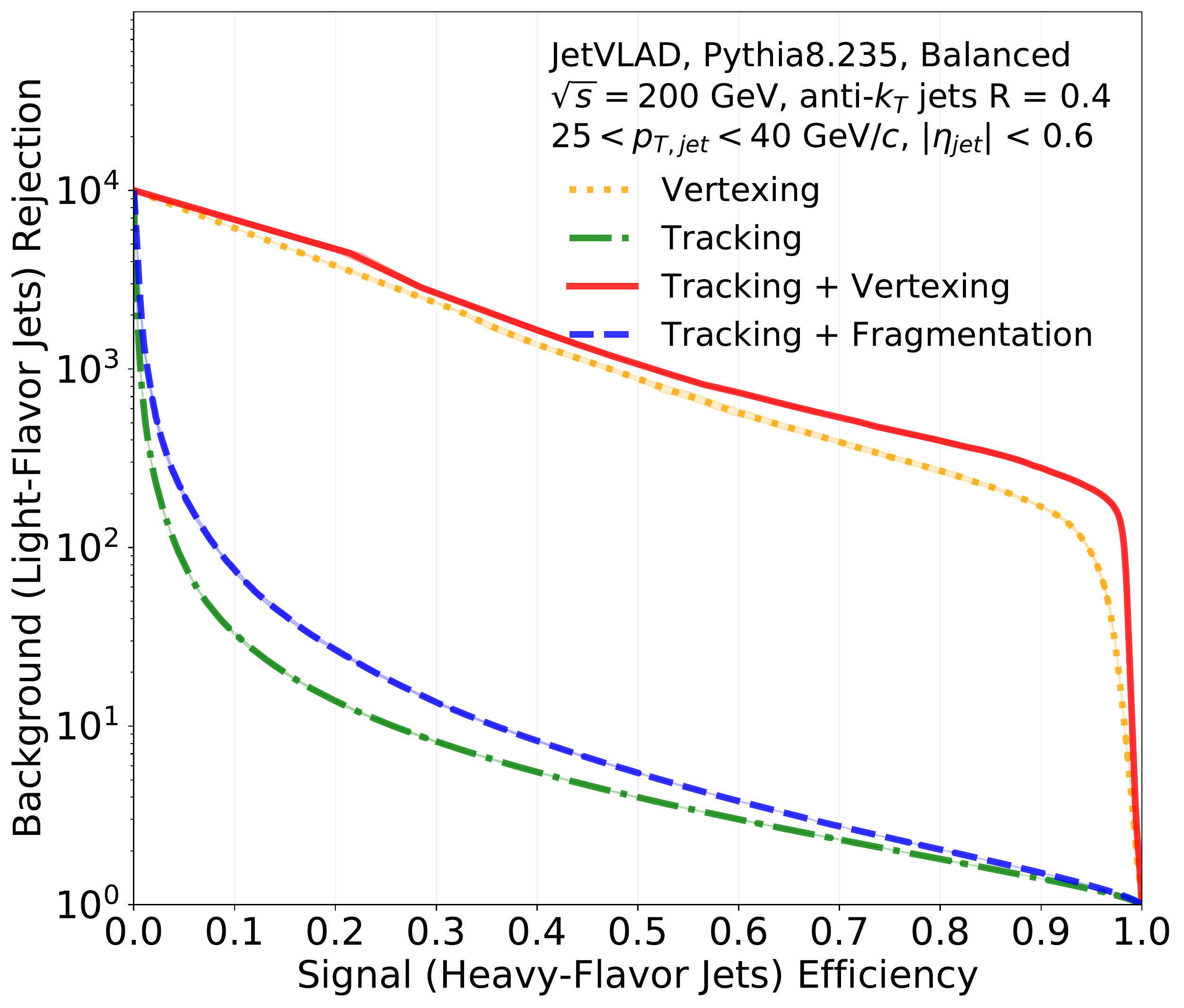}
    \caption{Background rejection vs efficiency curves shown for different inputs in the lines (described in the text) for balanced samples. The left and right panels show jets with $10 < p_{T} < 15$ and $25 < p_{T} < 40$ GeV/$c$.}
    \label{fig:results_balanced_rej_eff}
\end{figure}

Each sample of heavy-flavor jets, cross-section weighted and balanced, is trained and validated in parallel. The cross-section weighting is included in the datasets and is not considered explicitly in the training procedure. Once trained, we have two sets of model weights corresponding to the different datasets, which we can then use to further test the performance of \texttt{JetVLAD} on the respective samples. The background rejection vs efficiency curves for the balanced dataset are shown in Fig.\ref{fig:results_balanced_rej_eff}, where the left and right panels represent $R=0.4$ jets with $10< p_{T} < 15$ GeV/$c$ and $20 < p_{T} < 40$ GeV/$c$, respectively. 

\begin{figure}
    \centering
    \includegraphics[width=0.49\textwidth]{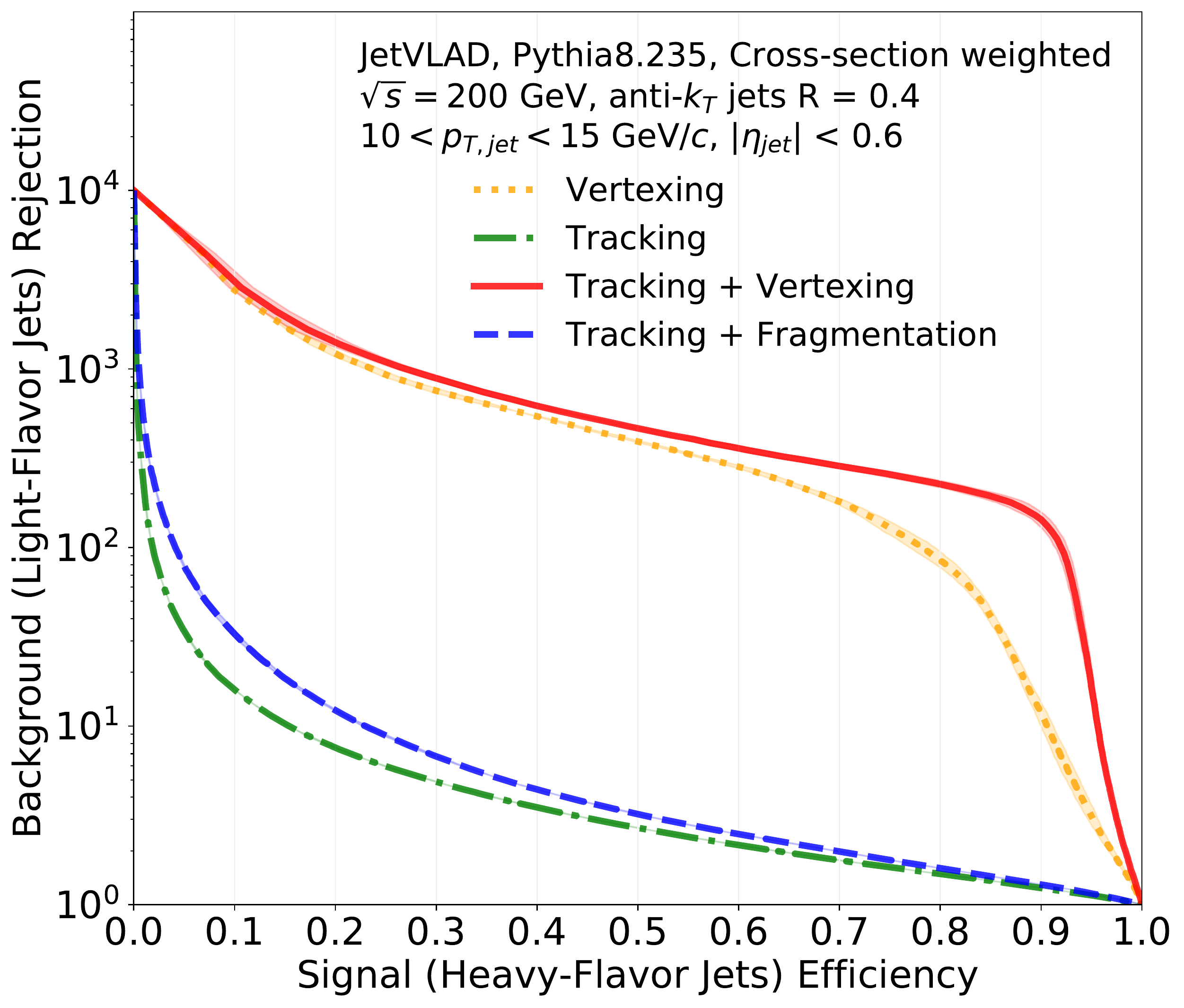}
    \includegraphics[width=0.49\textwidth]{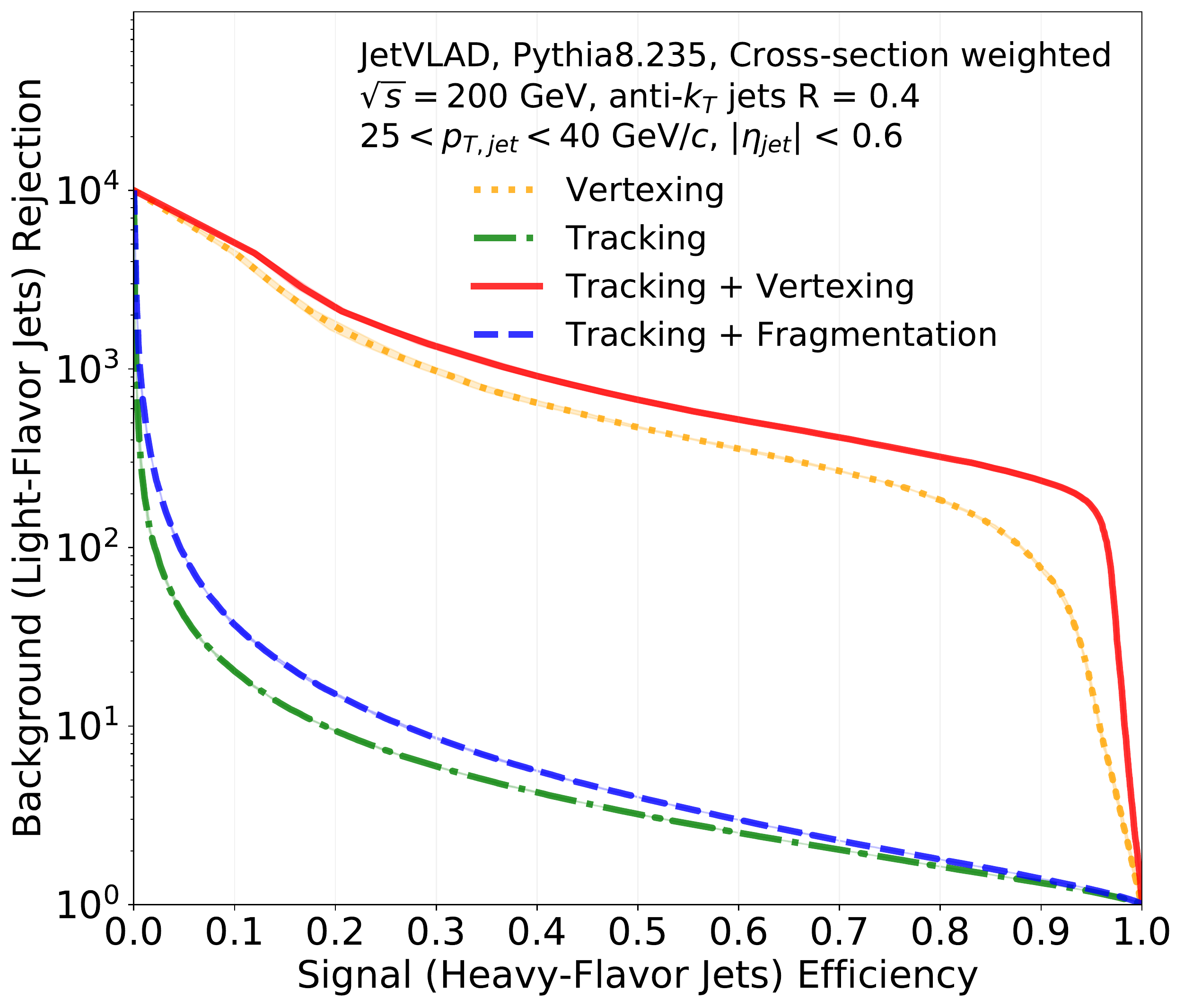}
    \caption{Background rejection vs efficiency curves shown for different inputs for cross-section weighted samples. The left and right panels show jets with $10 < p_{T} < 15$ and $25 < p_{T} < 40$ GeV/$c$.}
    \label{fig:results_hardqcd_rej_eff}
\end{figure}

Each curve in the plot represents different inputs to the model such as vertexing (dotted yellow), tracking (dot-dashed green), tracking $+$ vertexing (solid red) and tracking $+$ fragmentation (dashed blue). Similarly the performance for the cross-section weighted sample is shown in Fig.\ref{fig:results_hardqcd_rej_eff}. For both weighted and balanced datasets, we find the secondary vertex information provides the maximal impact on tagging heavy-flavor jets. The inclusion of tracking information in the input shows improved background rejection at large efficiencies while models trained with fragmentation information show a slight improvement on top of tracking at small efficiencies but do not significantly improve the performance at large efficiencies. 

\begin{figure}
    \centering
    \includegraphics[width=0.49\textwidth]{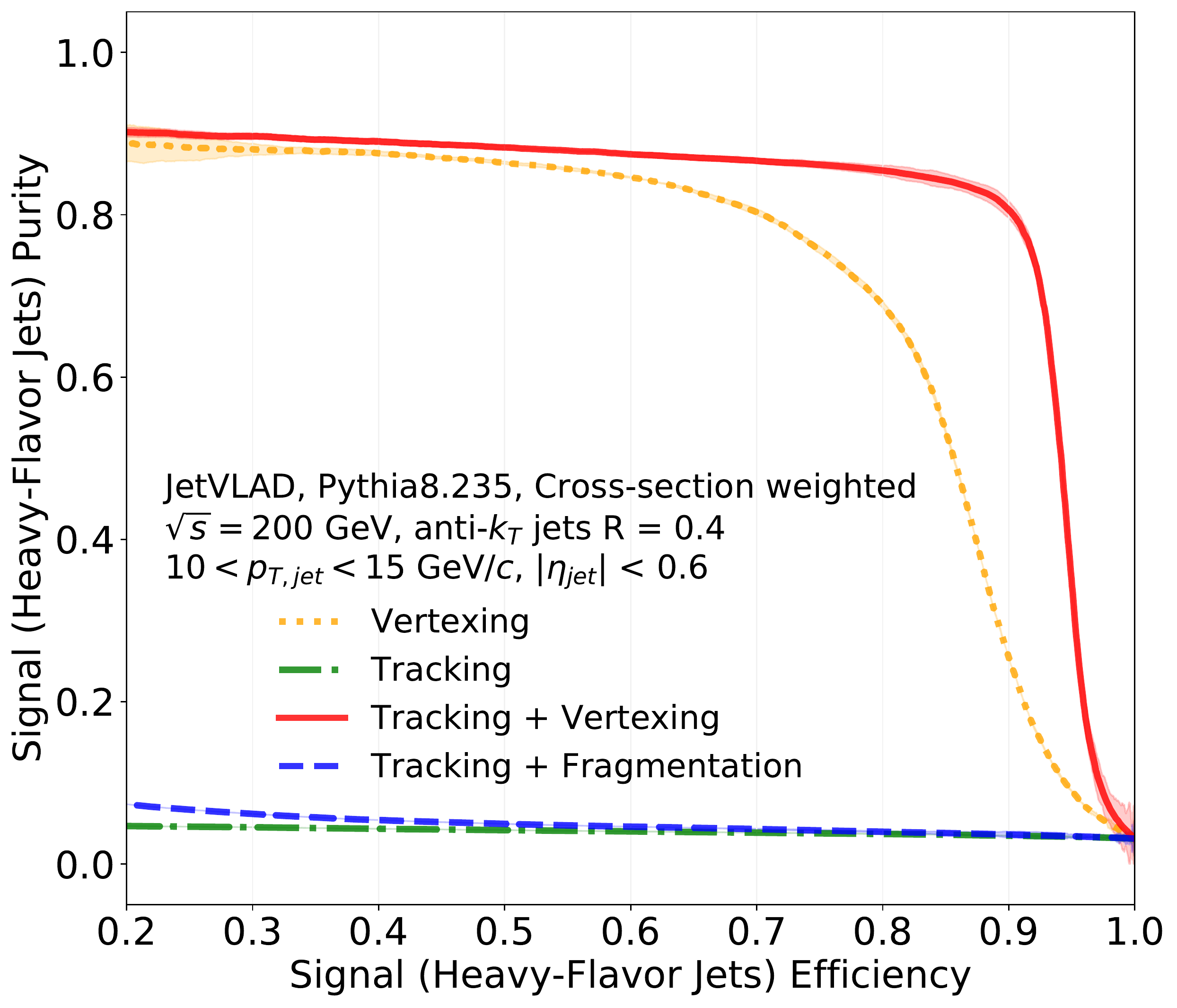}
    \includegraphics[width=0.49\textwidth]{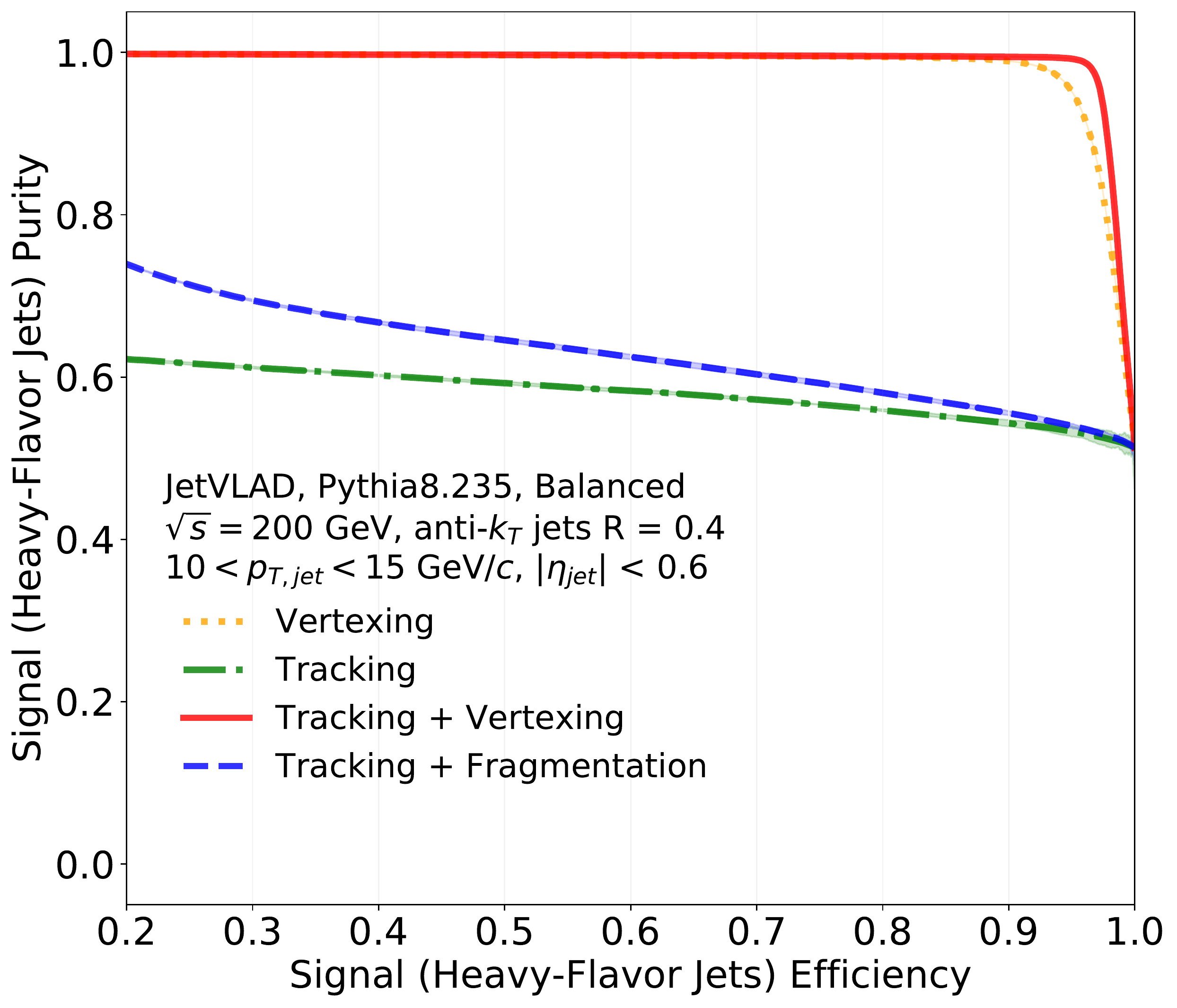}
    \includegraphics[width=0.49\textwidth]{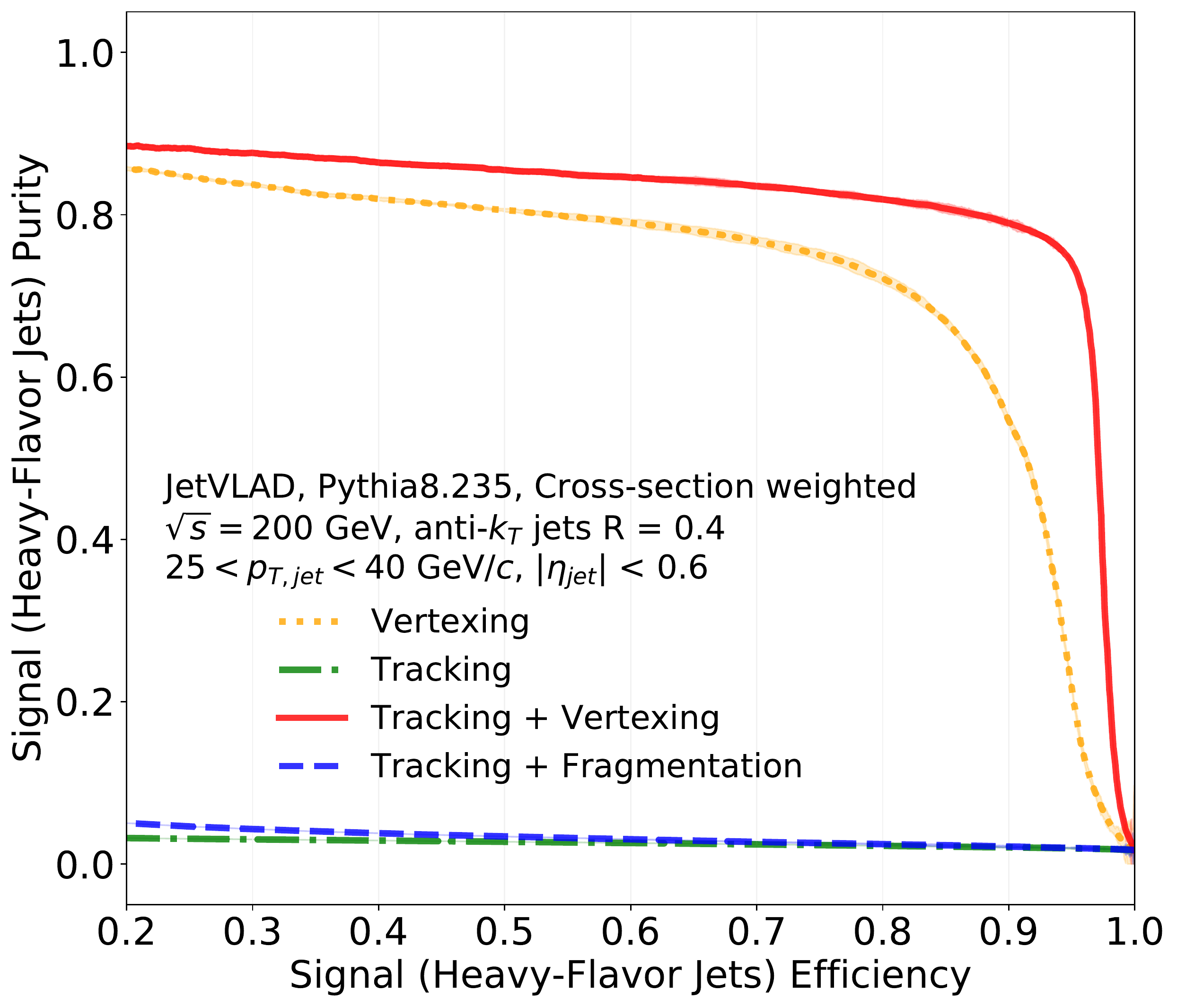}
    \includegraphics[width=0.49\textwidth]{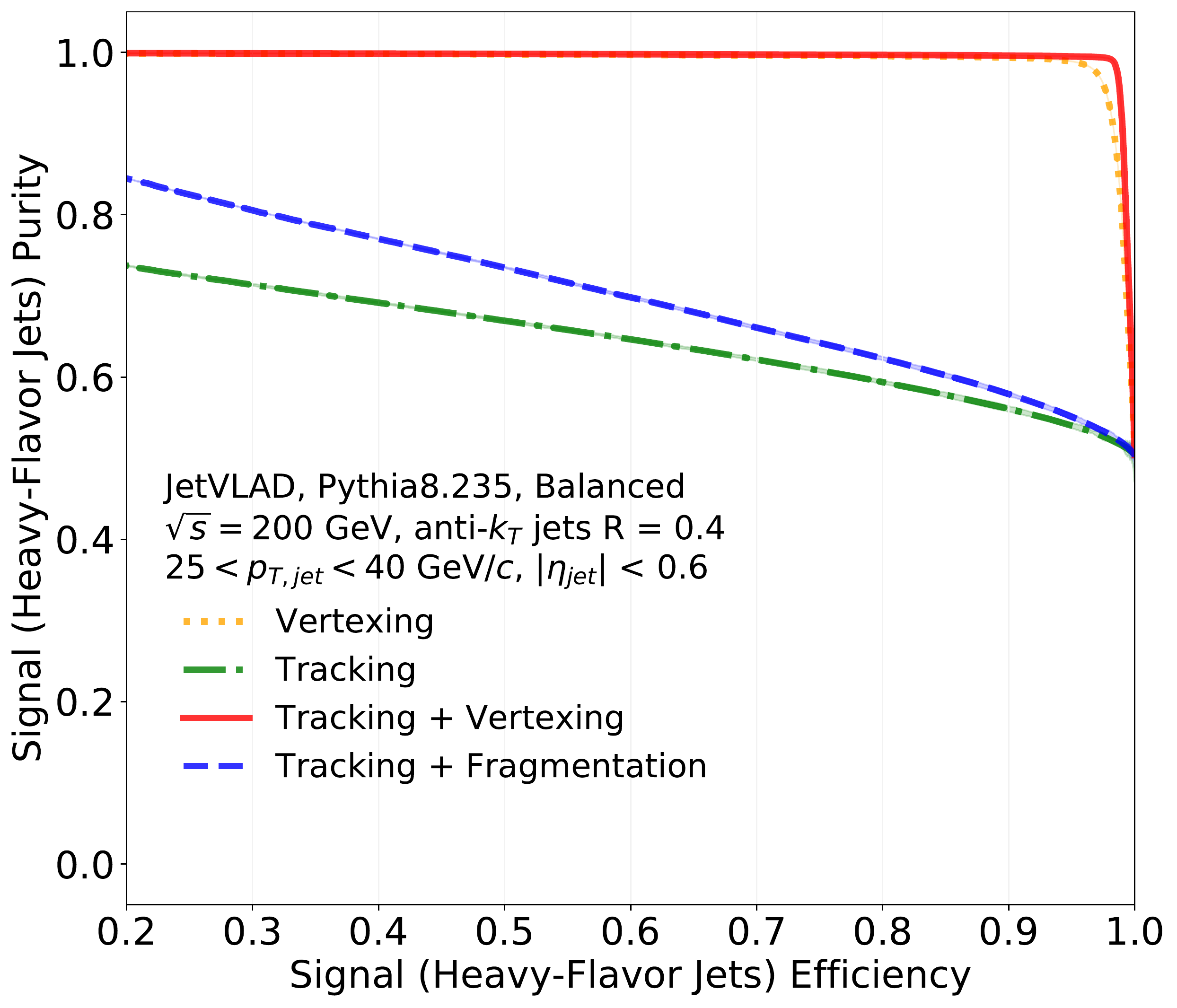}
    \caption{Purity vs efficiency curves shown for different inputs in the  colored lines and for cross-section weighted (left) and balanced (right) samples. The top and bottom panels show jets with $10 < p_{T} < 15$ and $25 < p_{T} < 40$ GeV/$c$.}
    \label{fig:results_purity_eff}
\end{figure}

We also study the purity vs efficiency for the different inputs and weighted samples in Fig.\ref{fig:results_purity_eff}. In contrast to the background rejection that did not show a large effect when considering the cross-section weighting, the purity on the other hand shows a distinct improvement for balanced samples. The purity for balanced dataset being close to $100\%$ can be understood due to the unrealistic yield of heavy-flavor jets in the sample. A more realistic performance, comparable to experimental data is shown for the cross-section weighted sample where for 80\% efficiency, we have close to $80\%$ purity and a background rejection of $236$. In both studies, we find the tracking $+$ vertexing still has the largest purity at given efficiency.  

\begin{figure}
    \centering
    \includegraphics[width=0.49\textwidth]{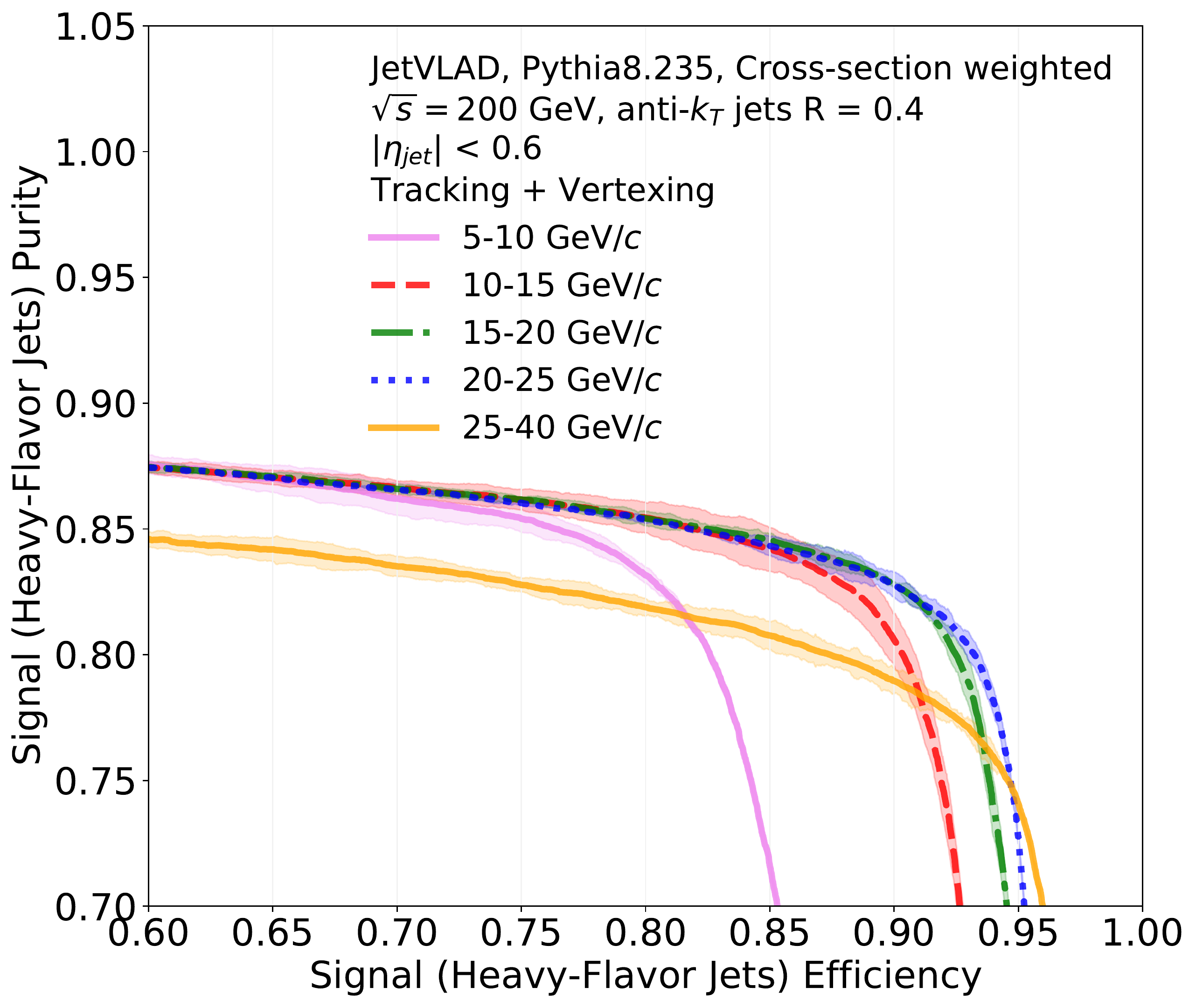}
    \includegraphics[width=0.49\textwidth]{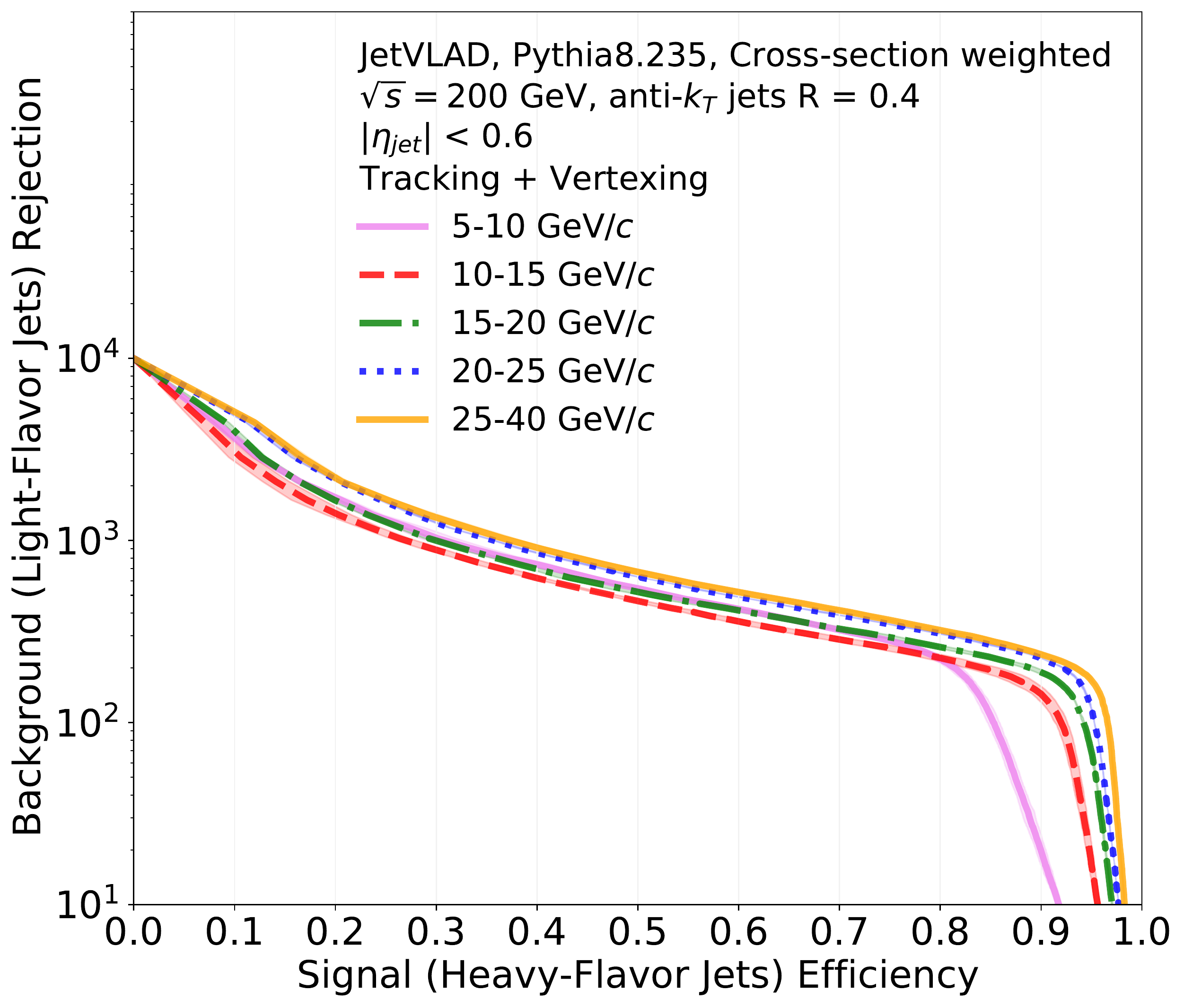}
    \caption{Purity (left) and background rejection (right) vs efficiency for tracking $+$ vertexing input. Each curve corresponds to a different jet $p_{T}$ selections.}
    \label{fig:ptcomp}
\end{figure}

The purity (left) and background rejection (right) as a function of jet momenta for the tracking $+$ vertexing input are shown in Fig.\ref{fig:ptcomp}. We find a consistent trend of increasing background rejections at given efficiencies with increasing jet $p_{T}$. At fixed purity of $70\%$, we also find a similar trend as before with increasing efficiency. At signal purity closer to $80\%$, we find an interesting trend for the highest momenta jets where the efficiency drops significantly indicating a kinematic effect of the jets and their substructure. We understand the drop in purity for the highest jet momenta to partly be due to the overlap in a high-level feature space between the light- and heavy-flavor samples and a further exploration of this behavior is reserved for an upcoming publication.   

We choose two working points based on efficiencies of $81\%$ and $50\%$ for which the corresponding purity and background rejections in cross-section weighted and balanced datasets are shown in Tab.~\ref{tab:rejecteff}. As we increase the jet $p_{T}$, we generally find that the signal purity is relatively consistent for both working points with a notable exception at the lowest and highest jet momenta. The background rejection increases in a non-linear fashion as jet momenta increase with the working point at $50\%$ efficiency having a rejection similar to the difference in the cross-sections. Extending the tagging to the lowest jet energies we see a drop in the purity at higher efficiencies due to the overlap between the jet topologies for light- and heavy-flavor jets where the reduced boost resulting in a smaller overall $DCA$. 
 
\begin{table}[h]
\begin{center}
\begin{tabular}{| c | c | c | c |}
\hline
Range in jet & Tagging & Signal & Background \\
$p_{\rm{T}}$ [GeV/$c$] & Efficiency & Purity & Rejection \\ [0.5ex]
\hline\hline
 [5 - 10]  & 80\% & 83\% ({\color{red} 99\%})  & 223 ({\color{red} 268}) \\ 
               & 50\% & 88\% ({\color{red} 99\%})  & 540 ({\color{red} 579})  \\ 
\hline
 [10 - 15] & 80\% & 85\% ({\color{red} 99\%})  & 223 ({\color{red} 230}) \\ 
               & 50\% & 88\% ({\color{red} 99\%})  & 476 ({\color{red} 449})  \\ 
\hline
 [15 - 20] & 80\% & 85\% ({\color{red} 99\%})  & 259 ({\color{red} 259}) \\ 
               & 50\% & 88\% ({\color{red} 99\%})  & 506 ({\color{red} 476})  \\ 
\hline
 [20 - 25] & 80\% & 85\% ({\color{red} 99\%})  & 310 ({\color{red} 336}) \\ 
               & 50\% & 88\% ({\color{red} 99\%})  & 624 ({\color{red} 740})  \\ 
\hline
 [25 - 40] & 80\% & 81\% ({\color{red} 99\%})  & 322 ({\color{red} 366}) \\ 
               & 50\% & 85\% ({\color{red} 99\%})  & 677 ({\color{red} 740})  \\ 
\hline
\end{tabular}
\caption{\texttt{JetVLAD} classification performance in purity and rejection for different jet $p_{T}$ ranges for the cross-section weighted ({\color{red} balanced}) datasets with two working points based on efficiencies of $81\%$ and $50\%$, respectively. Note: the balanced samples include unrealistic production cross-section for heavy flavor jets which results in artificially increased classification performance.}
\label{tab:rejecteff}
\end{center}
\end{table}

In our study, we applied a fast simulation of the STAR detector to compare the performance at the detector-level. In order for such a network to be fully applicable on data, one needs to also include additional effects such as out-of-time pileup for example. As the luminosity of the beams increases, the probability to have particles in your jet that arise from pileup vertices increases and these additional particles can affect the performance (since they tend to have very large $DCA$ values). The impact of out-of-time pile-up contribution to jets were studied by embedding minimum-bias \textsc{PYTHIA} events along with our hard-scattered event. The extent of pile-up contribution is dependent on the beam luminosities and data-taking rate and since they are relatively small at RHIC compared to the LHC, we run two scenarios of including 1 pile-up vertex, which is randomly placed along the z-direction of STAR's acceptance. We see an overall degradation of $\approx 2-3\%$ in the tagging purity at fixed efficiency (80\%). As the classification performances are non-monotonic to input variation, we recommend retraining the weights with an accurate simulation of pileup contribution to negate some of these degradation effects. In an upcoming publication we will focus more on the impact of in-time pileup such as the underlying event of heavy ion collisions and also increased pileup similar to what is observed at the LHC in order to stress-test the performance of the  \texttt{JetVLAD} tagger.

\section{Conclusions and Outlook}
\label{sec:conclusions}
We focused on identifying jets originating from heavy quarks such as $b$ and $c$, as opposed to those that originate from lighter quarks and gluons. We introduced the \texttt{JetVLAD} model which takes charged jet constituents with varying quantities as input and aggregates to a descriptor vector which can then be used to compare different jet populations. We trained the model on light- and heavy-flavor jets in \textsc{PYTHIA} and compared the classification performance for different varieties of track inputs based on metrics such as purity and background rejection at various signal efficiencies. In our studies we identified a combination of track inputs such as the secondary vertex ($DCA_{x,y}$, $DCA_{z}$) and the kinematics of the tracks ($p_{T}, \eta, \phi$) performed optimally leading to a signal purity of $85\%$ for an efficiency of $81\%$ when we consider the cross-section weighted sample. We also studied the effect of the jet momenta and found that with increasing jet momenta, we increased the background rejection while the signal purity was relatively consistent at a given efficiency which can be utilized as an experimental working point. Our studies highlight the importance of a precision vertex detector for heavy flavor studies. 

We demonstrated for the first time at RHIC energies the use of particle descriptors for identifying jet flavor. These low momenta jets at RHIC are particular important for studies of the QGP transport properties since they are the ones that are expected to have the largest interactions with the medium. Aggregated particle descriptors offer a feature space where the inherent differences between light- and heavy-flavor jets are highlighted leading to significant improvements in classification performance. 

Future explorations of this model include studying the effects of the heavy ion underlying event and also its extension to jet production at the LHC where one can explore BSM searches. While we have utilized the detector smearing based on the STAR Heavy-Flavor Tracker detector, this technique can be utilized to create a standard performance toolkit at sPHENIX~\cite{Adare:2015kwa} and in guiding detector design at the recently approved Electron Ion Collider (EIC). Studies at sPHENIX, whose vertex detector is designed for a high density heavy ion environment, are of immediate benefit for our model. Conversely, detectors at the EIC are more tuned towards precision measurements in a low density environment where one could take advantage of the \texttt{JetVlad} model architecture that effectively reduces the dimensionality of inputs to an aggregated vector, leading to variety of applications. In principle such a model can be utilized for studies related to particle flow, taking into account both charged and neutral candidates in experiment and also for mapping detector performance. 

\acknowledgments
We thank Dennis Perepelitsa, Leticia Cunqueiro and Marta Verweij for comments on the draft. We also like to thank Jan Kreps and the CIIRC IT department for computing help. Antoine Miech's LOUPE library~\cite{DBLP:journals/corr/MiechLS17} was used as the starting point for the \texttt{NetVLAD} layer. RKE thanks the Wayne State University computing grid for resources used in generating the training and testing datasets. RKE and JP are supported by US DOE grant No. DE-FG02-92ER40713. GP and JB are supported by project LTT18002 of the Ministry of Education, Youth and Sport of the Czech Republic. JB is supported by European Regional Development Fund-Project "Center of Advanced Applied Science" No. $CZ.02.1.01/0.0/0.0/16-019/0000778$. JS is partly supported by the the European Regional Development Fund under the project IMPACT (reg. no. $CZ.02.1.01/0.0/0.0/15\_003/0000468$) and by the French government under management of Agence Nationale de la Recherche as part of the "Investissements d'avenir" program, reference ANR-19-P3IA-0001 (PRAIRIE 3IA Institute).

\appendix
\section{Fast Simulation}
\label{app:fastsmear}
Here we describe the fast simulation framework that was used in order to simulate detector response. Since a full detector simulation via GEANT is time consuming, one can capture major effects via parametrizing the response. To account for the tracking efficiency at STAR in proton-proton collisions, each charged and final state track is dropped with probability of $20\%$~\cite{rusnakThesis}. The TPC also produces a momentum smearing~\cite{rusnakThesis}, which is modeled by 
\begin{equation}
    p_T = \mathcal{N}(p_T, 0.003 \cdot p_T^2).
\end{equation}
Regarding the finite vertex resolution, we apply a parametrization of the STAR heavy-flavor Tracker~\cite{}, by fitting the resolutions of $DCA_{xy}$ and $DCA_z$ dependent on the track momentum $\mathbf{p}$
\begin{equation}
    DCA_{xy} = \mathcal{N}(DCA_{xy}, \sigma_{xy}(\mathbf{p})),
\end{equation}
\begin{equation}
    DCA_{z} = \mathcal{N}(DCA_{z}, \sigma_{z}(\mathbf{p})).
\end{equation}

Post smearing, we apply selection criteria on the tracks similar to an experimental analysis 
\begin{itemize}
    \item Minimal smeared transverse momentum of the track is $p_T > 0.2$ GeV$/c$.
    \item $|DCA_z| < 60$ mm and $|DCA_{xy}| < 20$ mm
\end{itemize}

As mentioned before, this particular parametrization is based upon the existing STAR experiment. The upcoming  sPHENIX experiment is designed for better tracking efficiency and includes specific detector prioritizing secondary vertex resolution such as the MVTX~\cite{mvtxproposal} which will further increase the classification performance.

\bibliographystyle{JHEP.bst}
\bibliography{HFJetID_refs}

\providecommand{\href}[2]{#2}\begingroup\raggedright\begin{thebibliography}{10}

\bibitem{PhysRevLett.39.1436}
G.~Sterman and S.~Weinberg, {\it Jets from quantum chromodynamics},  {\em Phys.
  Rev. Lett.} {\bf 39} (Dec, 1977) 1436--1439.

\bibitem{PhysRevD.18.3320}
R.~P. Feynman, R.~D. Field, and G.~C. Fox, {\it Quantum-chromodynamic approach
  for the large-transverse-momentum production of particles and jets},  {\em
  Phys. Rev. D} {\bf 18} (Nov, 1978) 3320--3343.

\bibitem{Gribov:1972ri}
V.~N. Gribov and L.~N. Lipatov, {\it {Deep inelastic e p scattering in
  perturbation theory}},  {\em Sov. J. Nucl. Phys.} {\bf 15} (1972) 438--450.
  [Yad. Fiz.15,781(1972)].

\bibitem{Lipatov:1974qm}
L.~N. Lipatov, {\it {The parton model and perturbation theory}},  {\em Sov. J.
  Nucl. Phys.} {\bf 20} (1975) 94--102. [Yad. Fiz.20,181(1974)].

\bibitem{Dokshitzer:1977sg}
Y.~L. Dokshitzer, {\it {Calculation of the Structure Functions for Deep
  Inelastic Scattering and e+ e- Annihilation by Perturbation Theory in Quantum
  Chromodynamics.}},  {\em Sov. Phys. JETP} {\bf 46} (1977) 641--653. [Zh.
  Eksp. Teor. Fiz.73,1216(1977)].

\bibitem{Altarelli:1977zs}
G.~Altarelli and G.~Parisi, {\it {Asymptotic Freedom in Parton Language}},
  {\em Nucl. Phys.} {\bf B126} (1977) 298--318.

\bibitem{Catani:1996vz}
S.~Catani and M.~H. Seymour, {\it {A General algorithm for calculating jet
  cross-sections in NLO QCD}},  {\em Nucl. Phys.} {\bf B485} (1997) 291--419,
  [\href{http://arxiv.org/abs/hep-ph/9605323}{{\tt hep-ph/9605323}}]. [Erratum:
  Nucl. Phys.B510,503(1998)].

\bibitem{Andersson:1983ia}
B.~Andersson, G.~Gustafson, G.~Ingelman, and T.~Sjostrand, {\it {Parton
  Fragmentation and String Dynamics}},  {\em Phys. Rept.} {\bf 97} (1983)
  31--145.

\bibitem{Webber:1983if}
B.~R. Webber, {\it {A QCD Model for Jet Fragmentation Including Soft Gluon
  Interference}},  {\em Nucl. Phys.} {\bf B238} (1984) 492--528.

\bibitem{Wobisch:1998wt}
M.~Wobisch and T.~Wengler, {\it {Hadronization corrections to jet
  cross-sections in deep inelastic scattering}},  in {\em {Monte Carlo
  generators for HERA physics. Proceedings, Workshop, Hamburg, Germany,
  1998-1999}}, pp.~270--279, 1998.
\newblock \href{http://arxiv.org/abs/hep-ph/9907280}{{\tt hep-ph/9907280}}.

\bibitem{Currie:2016bfm}
J.~Currie, E.~W.~N. Glover, and J.~Pires, {\it {Next-to-Next-to Leading Order
  QCD Predictions for Single Jet Inclusive Production at the LHC}},  {\em Phys.
  Rev. Lett.} {\bf 118} (2017), no.~7 072002,
  [\href{http://arxiv.org/abs/1611.01460}{{\tt arXiv:1611.01460}}].

\bibitem{Kang:2017frl}
Z.-B. Kang, F.~Ringer, and I.~Vitev, {\it {Inclusive production of small radius
  jets in heavy-ion collisions}},  {\em Phys.\ Lett.\ B} {\bf 769} (2017)
  242--248, [\href{http://arxiv.org/abs/1701.05839}{{\tt arXiv:1701.05839}}].

\bibitem{Dasgupta:2018nvj}
M.~Dasgupta, F.~A. Dreyer, K.~Hamilton, P.~F. Monni, and G.~P. Salam, {\it
  {Logarithmic accuracy of parton showers: a fixed-order study}},  {\em JHEP}
  {\bf 09} (2018) 033, [\href{http://arxiv.org/abs/1805.09327}{{\tt
  arXiv:1805.09327}}].

\bibitem{Dasgupta:2020fwr}
M.~Dasgupta, F.~A. Dreyer, K.~Hamilton, P.~F. Monni, G.~P. Salam, and G.~Soyez,
  {\it {Parton showers beyond leading logarithmic accuracy}},
  \href{http://arxiv.org/abs/2002.11114}{{\tt arXiv:2002.11114}}.

\bibitem{Britzger:2017maj}
D.~Britzger, K.~Rabbertz, D.~Savoiu, G.~Sieber, and M.~Wobisch, {\it
  {Determination of the strong coupling constant using inclusive jet cross
  section data from multiple experiments}},  {\em Eur. Phys. J.} {\bf C79}
  (2019), no.~1 68, [\href{http://arxiv.org/abs/1712.00480}{{\tt
  arXiv:1712.00480}}].

\bibitem{Burke:2013yra}
{\bf JET} Collaboration, K.~M. Burke et~al., {\it {Extracting the jet transport
  coefficient from jet quenching in high-energy heavy-ion collisions}},  {\em
  Phys. Rev.} {\bf C90} (2014), no.~1 014909,
  [\href{http://arxiv.org/abs/1312.5003}{{\tt arXiv:1312.5003}}].

\bibitem{Soltz:2019aea}
{\bf Jetscape} Collaboration, R.~Soltz, {\it {Bayesian extraction of $\hat{q}$
  with multi-stage jet evolution approach}},  {\em PoS} {\bf HardProbes2018}
  (2019) 048.

\bibitem{Connors:2017ptx}
M.~Connors, C.~Nattrass, R.~Reed, and S.~Salur, {\it {Jet measurements in heavy
  ion physics}},  {\em Rev. Mod. Phys.} {\bf 90} (2018) 025005,
  [\href{http://arxiv.org/abs/1705.01974}{{\tt arXiv:1705.01974}}].

\bibitem{Qin:2015srf}
G.-Y. Qin and X.-N. Wang, {\it {Jet quenching in high-energy heavy-ion
  collisions}},  {\em Int. J. Mod. Phys.} {\bf E24} (2015), no.~11 1530014,
  [\href{http://arxiv.org/abs/1511.00790}{{\tt arXiv:1511.00790}}].
  [,309(2016)].

\bibitem{Blaizot:2015lma}
J.-P. Blaizot and Y.~Mehtar-Tani, {\it {Jet Structure in Heavy Ion
  Collisions}},  {\em Int. J. Mod. Phys.} {\bf E24} (2015), no.~11 1530012,
  [\href{http://arxiv.org/abs/1503.05958}{{\tt arXiv:1503.05958}}].

\bibitem{Dokshitzer_1991}
Y.~L. Dokshitzer, V.~A. Khoze, and S.~I. Troyan, {\it On specific qcd
  properties of heavy quark fragmentation (dead cone)},  {\em Journal of
  Physics G: Nuclear and Particle Physics} {\bf 17} (oct, 1991) 1602--1604.

\bibitem{Abdallah:2008ac}
{\bf DELPHI} Collaboration, J.~Abdallah et~al., {\it {Study of b-quark mass
  effects in multijet topologies with the DELPHI detector at LEP}},  {\em Eur.
  Phys. J.} {\bf C55} (2008) 525--538,
  [\href{http://arxiv.org/abs/0804.3883}{{\tt arXiv:0804.3883}}].

\bibitem{Maltoni:2016ays}
F.~Maltoni, M.~Selvaggi, and J.~Thaler, {\it {Exposing the dead cone effect
  with jet substructure techniques}},  {\em Phys. Rev.} {\bf D94} (2016), no.~5
  054015, [\href{http://arxiv.org/abs/1606.03449}{{\tt arXiv:1606.03449}}].

\bibitem{Cunqueiro:2018jbh}
L.~Cunqueiro and M.~Płoskoń, {\it {Searching for the dead cone effects with
  iterative declustering of heavy-flavor jets}},  {\em Phys. Rev.} {\bf D99}
  (2019), no.~7 074027, [\href{http://arxiv.org/abs/1812.00102}{{\tt
  arXiv:1812.00102}}].

\bibitem{Zardoshti:2020cwl}
{\bf ALICE} Collaboration, N.~Zardoshti, {\it {First Direct Observation of the
  Dead-Cone Effect}},  \href{http://arxiv.org/abs/2004.05968}{{\tt
  arXiv:2004.05968}}.

\bibitem{Chatrchyan:2013exa}
{\bf CMS} Collaboration, S.~Chatrchyan et~al., {\it {Evidence of b-Jet
  Quenching in PbPb Collisions at $\sqrt{s_{NN}}=2.76$ TeV}},  {\em Phys. Rev.
  Lett.} {\bf 113} (2014), no.~13 132301,
  [\href{http://arxiv.org/abs/1312.4198}{{\tt arXiv:1312.4198}}]. [Erratum:
  Phys. Rev. Lett.115,no.2,029903(2015)].

\bibitem{Sirunyan:2018jju}
{\bf CMS} Collaboration, A.~M. Sirunyan et~al., {\it {Comparing transverse
  momentum balance of b jet pairs in pp and PbPb collisions at $
  \sqrt{s_{\mathrm{NN}}}=5.02 $ TeV}},  {\em JHEP} {\bf 03} (2018) 181,
  [\href{http://arxiv.org/abs/1802.00707}{{\tt arXiv:1802.00707}}].

\bibitem{Kang:2018wrs}
Z.-B. Kang, J.~Reiten, I.~Vitev, and B.~Yoon, {\it {Light and heavy flavor
  dijet production and dijet mass modification in heavy ion collisions}},  {\em
  Phys. Rev.} {\bf D99} (2019), no.~3 034006,
  [\href{http://arxiv.org/abs/1810.10007}{{\tt arXiv:1810.10007}}].

\bibitem{Li:2018ybp}
H.~T. Li and I.~Vitev, {\it {Jet splitting function in the vacuum and QCD
  medium}},  {\em PoS} {\bf HardProbes2018} (2018) 077,
  [\href{http://arxiv.org/abs/1812.03348}{{\tt arXiv:1812.03348}}].

\bibitem{Aad:2015ydr}
{\bf ATLAS} Collaboration, G.~Aad et~al., {\it {Performance of $b$-Jet
  Identification in the ATLAS Experiment}},  {\em JINST} {\bf 11} (2016),
  no.~04 P04008, [\href{http://arxiv.org/abs/1512.01094}{{\tt
  arXiv:1512.01094}}].

\bibitem{Chatrchyan:2012jua}
{\bf CMS} Collaboration, S.~Chatrchyan et~al., {\it {Identification of b-Quark
  Jets with the CMS Experiment}},  {\em JINST} {\bf 8} (2013) P04013,
  [\href{http://arxiv.org/abs/1211.4462}{{\tt arXiv:1211.4462}}].

\bibitem{Ilten:2017rbd}
P.~Ilten, N.~L. Rodd, J.~Thaler, and M.~Williams, {\it {Disentangling Heavy
  Flavor at Colliders}},  {\em Phys. Rev.} {\bf D96} (2017), no.~5 054019,
  [\href{http://arxiv.org/abs/1702.02947}{{\tt arXiv:1702.02947}}].

\bibitem{Voutilainen:2015lqa}
M.~Voutilainen, {\it {Heavy quark jets at the LHC}},  {\em Int. J. Mod. Phys.}
  {\bf A30} (2015), no.~31 1546008,
  [\href{http://arxiv.org/abs/1509.05026}{{\tt arXiv:1509.05026}}].

\bibitem{hanseulOhSTARposter}
{\bf STAR Collaboration} Collaboration, S.~Oh, ``{Performance of Heavy-flavor
  Tagged Jet Identification in STAR}.'' 2018.

\bibitem{Jung:2016yzu}
K.~Jung, {\em {Flavors in the Soup: An Overview of Heavy-Flavored Jet Energy
  Loss at CMS}}.
\newblock PhD thesis, Illinois U., Chicago, 2016.

\bibitem{PhysRevD.94.112002}
D.~Guest, J.~Collado, P.~Baldi, S.-C. Hsu, G.~Urban, and D.~Whiteson, {\it Jet
  flavor classification in high-energy physics with deep neural networks},
  {\em Phys. Rev. D} {\bf 94} (Dec, 2016) 112002.

\bibitem{Sirunyan:2019wwa}
{\bf CMS} Collaboration, A.~M. Sirunyan et~al., {\it {A deep neural network for
  simultaneous estimation of b jet energy and resolution}},
  \href{http://arxiv.org/abs/1912.06046}{{\tt arXiv:1912.06046}}.

\bibitem{ATL-PHYS-PUB-2017-003}
{\bf ATLAS Collaboration} Collaboration, A.~Collaboration, {\it {Identification
  of Jets Containing $b$-Hadrons with Recurrent Neural Networks at the ATLAS
  Experiment}},  Tech. Rep. ATL-PHYS-PUB-2017-003, CERN, Geneva, Mar, 2017.

\bibitem{ATL-PHYS-PUB-2017-013}
{\bf ATLAS Collaboration} Collaboration, A.~Collaboration, {\it {Optimisation
  and performance studies of the ATLAS $b$-tagging algorithms for the 2017-18
  LHC run}},  Tech. Rep. ATL-PHYS-PUB-2017-013, CERN, Geneva, Jul, 2017.

\bibitem{Sjostrand:2014zea}
T.~Sjöstrand, S.~Ask, J.~R. Christiansen, R.~Corke, N.~Desai, P.~Ilten,
  S.~Mrenna, S.~Prestel, C.~O. Rasmussen, and P.~Z. Skands, {\it {An
  Introduction to PYTHIA 8.2}},  {\em Comput. Phys. Commun.} {\bf 191} (2015)
  159--177, [\href{http://arxiv.org/abs/1410.3012}{{\tt arXiv:1410.3012}}].

\bibitem{ALICE-PUBLIC-2020-002}
{\bf ALICE Collaboration} Collaboration, {\it {Groomed jet substructure
  measurements of charm jets tagged with $\rm{D}^{0}$ mesons in pp collisions
  at $\sqrt{s}$ = 13 TeV}}, .

\bibitem{star_detector}
{\bf STAR} Collaboration, K.~Ackermann et~al., {\it {STAR Detector overview}},
  {\em Nucl. Instrum. Meth.} {\bf A499} (2003) 624--632.

\bibitem{Anderson:2003ur}
M.~Anderson et~al., {\it {The Star time projection chamber: A Unique tool for
  studying high multiplicity events at RHIC}},  {\em Nucl. Instrum. Meth.} {\bf
  A499} (2003) 659--678, [\href{http://arxiv.org/abs/nucl-ex/0301015}{{\tt
  nucl-ex/0301015}}].

\bibitem{QIU20141141}
H.~Qiu, {\it Star heavy flavor tracker},  {\em Nuclear Physics A} {\bf 931}
  (2014) 1141 -- 1146. QUARK MATTER 2014.

\bibitem{Cacciari:2008gp}
M.~Cacciari, G.~P. Salam, and G.~Soyez, {\it {The Anti-k(t) jet clustering
  algorithm}},  {\em JHEP} {\bf 04} (2008) 063,
  [\href{http://arxiv.org/abs/0802.1189}{{\tt arXiv:0802.1189}}].

\bibitem{Cacciari:2011ma}
M.~Cacciari, G.~P. Salam, and G.~Soyez, {\it {FastJet User Manual}},  {\em
  Eur.\ Phys.\ J.\ C} {\bf 72} (2012) 1896,
  [\href{http://arxiv.org/abs/1111.6097}{{\tt arXiv:1111.6097}}].

\bibitem{Kang:2018qra}
Z.-B. Kang, K.~Lee, and F.~Ringer, {\it {Jet angularity measurements for single
  inclusive jet production}},  {\em JHEP} {\bf 04} (2018) 110,
  [\href{http://arxiv.org/abs/1801.00790}{{\tt arXiv:1801.00790}}].

\bibitem{PhysRevC.102.054913}
{\bf STAR Collaboration} Collaboration, J.~Adam et~al., {\it Measurement of
  inclusive charged-particle jet production in au $+$ au collisions at
  $\sqrt{{s}_{NN}}=200$ gev},  {\em Phys. Rev. C} {\bf 102} (Nov, 2020) 054913.

\bibitem{Adam:2020kug}
{\bf STAR} Collaboration, J.~Adam et~al., {\it {Measurement of groomed jet
  substructure observables in p+p collisions at $\sqrt {s}$ =200 GeV with
  STAR}},  {\em Phys. Lett. B} {\bf 811} (2020) 135846,
  [\href{http://arxiv.org/abs/2003.02114}{{\tt arXiv:2003.02114}}].

\bibitem{Larkoski:2017jix}
A.~J. Larkoski, I.~Moult, and B.~Nachman, {\it {Jet Substructure at the Large
  Hadron Collider: A Review of Recent Advances in Theory and Machine
  Learning}},  {\em Phys. Rept.} {\bf 841} (2020) 1--63,
  [\href{http://arxiv.org/abs/1709.04464}{{\tt arXiv:1709.04464}}].

\bibitem{Catani:1993hr}
S.~Catani, Y.~L. Dokshitzer, M.~H. Seymour, and B.~R. Webber, {\it
  {Longitudinally invariant $K_t$ clustering algorithms for hadron hadron
  collisions}},  {\em Nucl. Phys.} {\bf B406} (1993) 187--224.

\bibitem{Dokshitzer:1997in}
Y.~L. Dokshitzer, G.~D. Leder, S.~Moretti, and B.~R. Webber, {\it {Better jet
  clustering algorithms}},  {\em JHEP} {\bf 08} (1997) 001,
  [\href{http://arxiv.org/abs/hep-ph/9707323}{{\tt hep-ph/9707323}}].

\bibitem{Andreassen:2018apy}
A.~Andreassen, I.~Feige, C.~Frye, and M.~D. Schwartz, {\it {JUNIPR: a Framework
  for Unsupervised Machine Learning in Particle Physics}},  {\em Eur. Phys. J.}
  {\bf C79} (2019), no.~2 102, [\href{http://arxiv.org/abs/1804.09720}{{\tt
  arXiv:1804.09720}}].

\bibitem{Arandjelovic18}
R.~{Arandjelović}, P.~{Gronat}, A.~{Torii}, T.~{Pajdla}, and J.~{Sivic}, {\it
  Netvlad: Cnn architecture for weakly supervised place recognition},  {\em
  IEEE Transactions on Pattern Analysis and Machine Intelligence} {\bf 40}
  (2018), no.~6 1437--1451.

\bibitem{DBLP:journals/corr/MiechLS17}
A.~Miech, I.~Laptev, and J.~Sivic, {\it Learnable pooling with context gating
  for video classification},  {\em CoRR} {\bf abs/1706.06905} (2017)
  [\href{http://arxiv.org/abs/1706.06905}{{\tt arXiv:1706.06905}}].

\bibitem{DBLP:journals/corr/HeZRS15}
K.~He, X.~Zhang, S.~Ren, and J.~Sun, {\it Deep residual learning for image
  recognition},  {\em CoRR} {\bf abs/1512.03385} (2015)
  [\href{http://arxiv.org/abs/1512.03385}{{\tt arXiv:1512.03385}}].

\bibitem{10.5555/2627435.2670313}
N.~Srivastava, G.~Hinton, A.~Krizhevsky, I.~Sutskever, and R.~Salakhutdinov,
  {\it Dropout: A simple way to prevent neural networks from overfitting},
  {\em J. Mach. Learn. Res.} {\bf 15} (Jan., 2014) 1929–1958.

\bibitem{DBLP:journals/corr/LoshchilovH16a}
I.~Loshchilov and F.~Hutter, {\it {SGDR:} stochastic gradient descent with
  restarts},  {\em CoRR} {\bf abs/1608.03983} (2016)
  [\href{http://arxiv.org/abs/1608.03983}{{\tt arXiv:1608.03983}}].

\bibitem{DBLP:journals/corr/abs-1907-06890}
M.~Seg{\`{u}}, A.~Loquercio, and D.~Scaramuzza, {\it A general framework for
  uncertainty estimation in deep learning},  {\em CoRR} {\bf abs/1907.06890}
  (2019) [\href{http://arxiv.org/abs/1907.06890}{{\tt arXiv:1907.06890}}].

\bibitem{Englert:2018cfo}
C.~Englert, P.~Galler, P.~Harris, and M.~Spannowsky, {\it {Machine Learning
  Uncertainties with Adversarial Neural Networks}},  {\em Eur. Phys. J.} {\bf
  C79} (2019), no.~1 4, [\href{http://arxiv.org/abs/1807.08763}{{\tt
  arXiv:1807.08763}}].

\bibitem{Kasieczka:2020vlh}
G.~Kasieczka, M.~Luchmann, F.~Otterpohl, and T.~Plehn, {\it {Per-Object
  Systematics using Deep-Learned Calibration}},
  \href{http://arxiv.org/abs/2003.11099}{{\tt arXiv:2003.11099}}.

\bibitem{Adare:2015kwa}
{\bf PHENIX} Collaboration, A.~Adare et~al., {\it {An Upgrade Proposal from the
  PHENIX Collaboration}},  \href{http://arxiv.org/abs/1501.06197}{{\tt
  arXiv:1501.06197}}.

\bibitem{rusnakThesis}
J.~Rusnak, {\em {Jet Reconstruction in Au+Au collisions at RHIC}}.
\newblock PhD thesis, Nuclear Physics Institute, The Czech Academy of Sciences,
  2017.

\bibitem{mvtxproposal}
``A monolithic active pixel sensor detector for the sphenix experiment.''
  \url{https://p25ext.lanl.gov/maps/mvtx/Proposals/sPHENIX-MVTX-Preproposal-022017-final.pdf}.
\newblock Feb 2017.

\end{thebibliography}\endgroup

\end{document}